\documentclass{aa}  
\usepackage{graphicx}
\usepackage{txfonts}
\usepackage{multirow}
\usepackage{booktabs}
\usepackage{xcolor}
\usepackage{natbib}
\usepackage{amsmath}
\usepackage[colorlinks=true,citecolor=blue,breaklinks]{hyperref}
\usepackage{caption}
\usepackage{subcaption}

\def\ms{\hbox{\,m\,s$^{-1}$}}         

\begin{document}

   \title{TOI-179: a young system with a transiting compact Neptune-mass planet and a low-mass companion in outer orbit
\thanks{Based on observations made with ESO Telescopes at the La Silla Paranal Observatory under programmes ID 
192.C-0224(C), 
098.C-0739(A), 
0101.C-0510(D), 
0102.C-0451(B), 
0102.D-0483(A), 
0102.C-0525(A) 
0103.C-0759(A) (HARPS), 
0104.C-0247(B) (SPHERE), 
097.C-0972(A) (NACO) 
} 
\fnmsep\thanks{This paper includes data collected by the TESS mission, which are publicly available from the Mikulski Archive for Space Telescopes (MAST).}}

\author{S. Desidera,\inst{1} 
M. Damasso,\inst{2}
R. Gratton,\inst{1}
S. Benatti,\inst{3}
D. Nardiello,\inst{1,4}
V. D'Orazi,\inst{1,5}
A.F. Lanza,\inst{6}
D. Locci,\inst{3}
F. Marzari,\inst{7,1}
D. Mesa,\inst{1}
S. Messina,\inst{4} 
I. Pillitteri,\inst{3}
A. Sozzetti,\inst{2}
J. Girard,\inst{8}
A. Maggio,\inst{3}
G. Micela,\inst{3}
L. Malavolta,\inst{9,1}
V. Nascimbeni,\inst{1,9}
M. Pinamonti,\inst{2}
V. Squicciarini,\inst{1,9}
J. Alcal\'a,\inst{10}
K. Biazzo,\inst{11}
A. Bohn,\inst{12}
M. Bonavita,\inst{13,14,1}
K. Brooks,\inst{15}
G. Chauvin,\inst{16}
E. Covino,\inst{10}
P. Delorme,\inst{17}
J. Hagelberg,\inst{18}
M. Janson,\inst{19}
A.-M. Lagrange,\inst{17}
C. Lazzoni,\inst{20,1}
}

\institute{ INAF -- Osservatorio Astronomico di Padova, Vicolo dell'Osservatorio 5, I-35122, Padova, Italy  
\and INAF -- Osservatorio Astrofisico di Torino, Via Osservatorio 20, I-10025, Pino Torinese (TO), Italy 
\and INAF -- Osservatorio Astronomico di Palermo, Piazza del Parlamento, 1, I-90134, Palermo, Italy 
\and Aix Marseille Univ, CNRS, CNES, LAM, Marseille, France 
\and Monash Centre for Astrophysics, School of Physics and Astronomy, Monash University, Clayton 3800, Melbourne, Australia  
\and INAF -- Osservatorio Astrofisico di Catania, Via S. Sofia 78, I-95123, Catania, Italy 
\and Dipartimento di Fisica e Astronomia -- Universt\`a di Padova, Via Marzolo 8, I-35121 Padova, Italy  
\and Space Telescope Science Institute, 3700 San Martin Drive, Baltimore, MD 21218, USA 
\and Dipartimento di Fisica e Astronomia -- Universt\`a di Padova, Vicolo dell'Osservatorio 3, I-35122 Padova, Italy 
\and INAF -- Osservatorio Astronomico di Capodimonte, Salita Moiariello 16, I-80131, Napoli, Italy 
\and INAF -- Osservatorio Astronomico di Roma, Via Frascati 33, 00078 -- Monte Porzio Catone (Roma), Italy 
\and Leiden Observatory, Leiden University, PO Box 9513, 2300 RA Leiden, The Netherlands 
\and School of Physical Sciences, The Open University, Walton Hall, Milton Keynes, MK7 6AA 
\and SUPA, Institute for Astronomy, University of Edinburgh, Blackford Hill, Edinburgh EH9 3HJ, UK 
\and Laboratory for Atmospheric and Space Physics 1234 Innovation Dr. Boulder, CO 80303, USA 
\and Universit\'e C\^ote d’Azur, Observatoire de la C\^ote d’Azur, CNRS,
Laboratoire Lagrange, Bd de l’Observatoire, CS 34229, 06304 Nice
cedex 4, France 
\and Univ. Grenoble Alpes, CNRS, IPAG, F-38000 Grenoble, France 
\and Observatoire de Gen\`eve, Universit\'e de Gen\`eve, 51 ch. des Maillettes, CH-1290 Sauverny, Switzerland 
\and Department of Astronomy, Stockholm University, SE-106 91 Stockholm, Sweden  
\and University of Exeter, Astrophysics Group, Physics Building, Stocker Road, Exeter, EX4 4QL, UK. 
}

   \date{Received ; accepted }

 
  \abstract
{Transiting planets around young stars are key benchmarks for our understanding of planetary systems. One of such candidates was identified
around the K dwarf HD 18599 by TESS, labeled as TOI-179.} 
{We present the confirmation of the transiting planet and the characterization of the host star and of the TOI-179 system over a broad range of angular separations.}
{To this aim, we exploited the TESS photometric time series, intensive radial velocity monitoring performed with HARPS, and deep high-contrast imaging observations obtained with SPHERE and NACO at VLT. The inclusion of Gaussian processes regression analysis is effective to properly model the magnetic activity of the star and identify the Keplerian signature of the transiting planet.}
{The star, with an age of 400$\pm$100 Myr, is orbited by a transiting planet with 
period 4.137436 days, mass 24$\pm$7 
$M_{\oplus}$,  radius 2.62$^{+0.15}_{-0.12} R_{\oplus}$,
and significant eccentricity (0.34$^{+0.07}_{-0.09}$).
Adaptive optics observations identified a low-mass companion at the boundary between
brown dwarfs and very low mass stars (mass derived from luminosity 83$^{+4}_{-6} M_{J}$) at a very small projected separation (84.5  mas, 3.3 au at the distance of the star).
Coupling the imaging detection with the long-term radial velocity trend and the astrometric signature, we constrained the orbit
of the low mass companion, identifying two families of possible orbital solutions. }
{The TOI-179 system represents a high-merit laboratory for our understanding of the physical evolution of planets and other low-mass objects and of how the planet properties are influenced by dynamical effects and interactions with the parent star.}

   \keywords{exoplanets -- techniques: photometric, spectroscopic, radial velocity, imaging spectroscopy -- stars: individual: TOI-179
               }

\titlerunning{TOI-179}
\authorrunning{Desidera et al.}
\maketitle


\section{Introduction}
\label{sec:intro}

The observed architecture of a planetary system takes shape after significant evolution of the system itself has been occurring over time, with several mechanisms acting on different timescales: migration within the native disk \citep{lin1996}, expected to occur in a few Myrs before disk dissipation; planet-planet dynamical instabilities \citep{marzari1996}; gravitational interactions with passing bodies \citep[more frequent for stars in clusters, ][]{bonnell2001} and bound companions on wide orbits \citep[through, e.g., the Kozai mechanisms, ][]{wu2007}, and circularization of the orbit by  tides from the host stars \citep{rasio1996}, which could be active on much longer timescales. 
Understanding the original configurations of the systems and the timescales on which these various mechanisms work is easier when observing planetary systems at young ages, with planets closer to their formation time epoch and possibly also to their birthsites. Young ages also offer the unique opportunity of accessing the range of wide separations through direct imaging with sensitivity into the planetary regime, as young planets are brighter \citep{chauvin2018}.
Furthermore, the outer planets in planet-planet scattering events are expected to be lost with time and then more easily identified close to the parent stars at young ages \citep{veras2009}. Excess of companions at wide separation were identified for some classes of planetary systems \citep[e.g.,][]{bryan2019,fontanive2019}. 

Transit space missions are deeply changing our view of planetary systems at close separations. A large fraction of  (old) G, K, and M stars were found to host low-mass planets \citep[e.g., ][]{howard2012,bryson2021}. These planets show a large variety of system architectures and internal structures as inferred by bulk density measurements. 
However, most targets of the first missions CoRoT and Kepler are very distant ($\ge$0.5 kpc). The large distance and faintness make them prohibitive targets for direct imaging characterization, limiting searches to wide stellar companions \citep[e.g., ][]{horch2014,hirsch2017}. Also, the large distances, the confined observations to well-defined fields, and the faint magnitude of the stars represent major challenges for a proper age estimation of these targets, exploiting spectroscopic diagnostics, isochrone fitting, or kinematics. Instead, the Transiting Exoplanet Survey Satellite (TESS) mission \citep{tess} offers a much better perspective, as it covers nearly the whole sky and targets closer and brighter objects than those observed with Kepler. Therefore, nearby young stars with well characterized ages, members of moving groups, star clusters or even isolated field stars, are being searched for planetary transits. Previously, Kepler/K2 also monitored several fields centered on star forming regions and young clusters \citep{rizzuto2017}.

Transiting planets at young ages are key benchmarks for our understanding of planetary systems:
{\it (a)} Their frequency at various ages allows one to constrain the timescales of the migration mechanisms.
{\it (b)} Their radii are expected to be expanded with respect to those of older planets of similar masses \citep{linder2019}.
{\it (c)} Their expected low density and the large scale height make them prime targets for atmospheric characterization. Planetary evaporation processes are expected to be stronger in these cases, for the planets close enough to their parent stars. Also, for late-type host stars, the impact of very high levels of stellar magnetic activity on the planetary atmospheres can be studied.
{\it (d)} For giant planets, their structure and atmospheric properties can be compared to those of non-irradiated planets with similar age at wide separation, detected and characterized with the imaging technique.
{\it (e)} The transit geometry itself and the Rossiter-McLaughlin effect 
provide additional constraints on the architecture of the system, not available for planets detected only with the radial velocity (RV) technique.

Promising results have emerged from several efforts over the past few years.
These include the discovery of young multiplanet systems such as V1298 Tau \citep{david2019} and AU Mic \citep{plavchan2020}, with
determination of the masses of some planets in these systems obtained through intensive RV monitoring campaigns \citep{suarezmascareno2022,klein2021,zicher2022}.
Several other young transiting planets were identified around
stars which are members of open clusters \citep[e.g., ][]{bouma2020}, stellar associations and moving groups \citep[e.g.,][]{rizzuto2020,newton2021} or field objects \citep[e.g.][]{zhou2021,barragan2022}; in many cases these planets still lack reliable mass determination, as the observational efforts to overcome the noise linked to stellar activity are large for such active stars. 
Additional characterization studies were performed
in a few cases such as the measurement of Rossiter-McLaughlin effect
\citep[e.g.][]{mann2020}.

Our team started to work on TESS datasets since the first sector of
TESS observations, which included a planet candidate around DS Tuc, a bona-fide member of the Tuc-Hor association (age 40-45 Myr).
The transiting planet candidate detected by TESS was validated independently by \citet{benatti2019} and \citet{newton2019} and further characterized by \citet{benatti2021}. Northern young targets are being followed-up in the framework of the GAPS program \citep{carleo2020}, starting from the validation of the 
two-planet system around TOI-942, a 50 Myr old K star \citep{carleo2021} and the measurement of the mass of the ultra-short-period planet TOI-1807b around a 300 Myr old star \citep{nardiello2022}.

As part of on-going efforts to validate and characterise young
transiting exoplanets identified by TESS, we present in this paper
our analysis for the system observed around the star \object{HD 18599} = HIP 13754, a bright (V=8.99 mag)
and active K dwarf, also known as TESS Object of Interest (TOI)-179. The planet candidate was identified from TESS observations gathered in Sectors 2--3. The star was also reobserved by TESS during Sectors 29--30. It is a  single-planet candidate, with radius of 2.6 $R_{\oplus}$, and period of 4.1375 d, according to the TOI release information\footnote{ \url{https://tess.mit.edu/toi-releases/}  }. It was the target  of our dedicated investigations, using HARPS RV monitoring, and high contrast imaging observations with SPHERE and NaCo. 

The paper is organized as follows: 
Sect. \ref{sec:obs} presents the photometric, spectroscopic, and high-contrast imaging observations,
Sect. \ref{sec:star} summarizes the stellar parameters (whose determination is described in detail in
Appendix \ref{app:star}), 
Sect. \ref{sec:toi179b} presents the confirmation and mass determination of the transiting planet TOI-179b,
Sect. \ref{sec:bd} reports the detection of a low-mass companion at the BD-star boundary.
In Sect. \ref{sec:discussion}, we discuss the various implications of our findings
and finally in Sect. \ref{sec:summary} we summarize our results and the perspective for further characterization.

\section{Observations and data reduction}
\label{sec:obs}

\subsection{Photometric observations with TESS}
\label{sec:tess}

In this work we made use both of the {\it TESS} long-cadence (30-/10-minute) and short-cadence (2-minute) observations. Long-cadence data will be used to perform a series of vetting tests and to confirm the planetary nature of TOI-179~b (Sect.~\ref{sec:phot_toi179}), and to measure the rotation period (App. \ref{sec:rot}), while short-cadence observations will be adopted for the modeling of the planetary transits (Sect.~\ref{sec:rv_tess_fit_toi179}).

\subsubsection{Long cadence light curves from Full Frame Images}

 We used the Full Frame Images (FFIs) collected by {\it TESS} during the Cycles 1 and 3 to extract the light curves of TOI-179. This object was observed in 2018, between August 22nd and October 18th  in 30-minute cadence mode (Sectors 2 and 3), and again in 2020, between August 26th and October 21st
  with a cadence of 10-minutes (Sectors 29 and 30). 
 For the extraction of the light curves we used the light curve extractor developed by \citet{2015MNRAS.447.3536N,2016MNRAS.455.2337N} and adapted to {\it TESS} data by \citet{2019MNRAS.490.3806N}. Briefly, given a target star, the software subtracts from the FFIs all its neighbors by using information from an input catalog (Gaia~DR2) and empirical Point Spread Functions (PSFs). After the subtraction, it extracts the flux of the target star by using 5 different photometric apertures. The systematic effects that affect the light curves are finally corrected by fitting to them the Cotrending Basis Vectors extracted as described in \citet{2020MNRAS.495.4924N}.

\subsubsection{Short cadence light curves}

TOI-179 was observed by {\it TESS} in two minute cadence mode in Sectors 2, 3, 29, and 30  via target pixel stamp observations, and it was included in the following TESS Guest Investigator Programs: GO3102 (PI: T.~Oswalt), GO3227 (PI: J.~Davenport), GO3272 (PI: J.~Burt), and GO3278 (PI: A.~Mayo). In this work we used the Pre-search Data Conditioning Simple Aperture Photometry (PDCSAP) light curves (\citealt{2012PASP..124.1000S,2012PASP..124..985S,2014PASP..126..100S}) from the Science Processing Observation Center (SPOC, \citealt{2016SPIE.9913E..3EJ}), extracted from the two minute target pixel files. 




\subsection{Spectroscopic observations with HARPS}
\label{sec:harps}

We obtained high-resolution spectra of TOI-179 within a dedicated follow-up program with HARPS \citep{harps} at the ESO 3.6m telescope (La Silla Observatory, Chile, Program ID 0103.C-0759(A), PI Benatti). From Jul 6th to Sept 22nd, 2019, we collected 40 spectra of TOI-179 with typical exposure time of 900 s and signal-to-noise (S/N) of about 60. Additional spectra were collected under different ESO programs (192.C-0224 and  098.C-0739, PI Lagrange; 0101.C-0510, PI Jordan; 0102.C-0451 PI Brahm; 0102.D-0483, PI Berdinas; and 0102.C-0525, PI Diaz) and retrieved from the ESO archive. Part of the early dataset was published independently by \citet{grandjean2020}, with the RVs derived through the SAFIR software. Overall, in this work we analysed 103 radial velocity measurements of TOI-179, with a time baseline of 1768 d (from 2014 Nov 19 to 2019 Sept 22).  

\begin{figure}
    \centering
    \includegraphics[trim={0 1.5cm 0 0},clip, width=0.39\textwidth, angle=270]{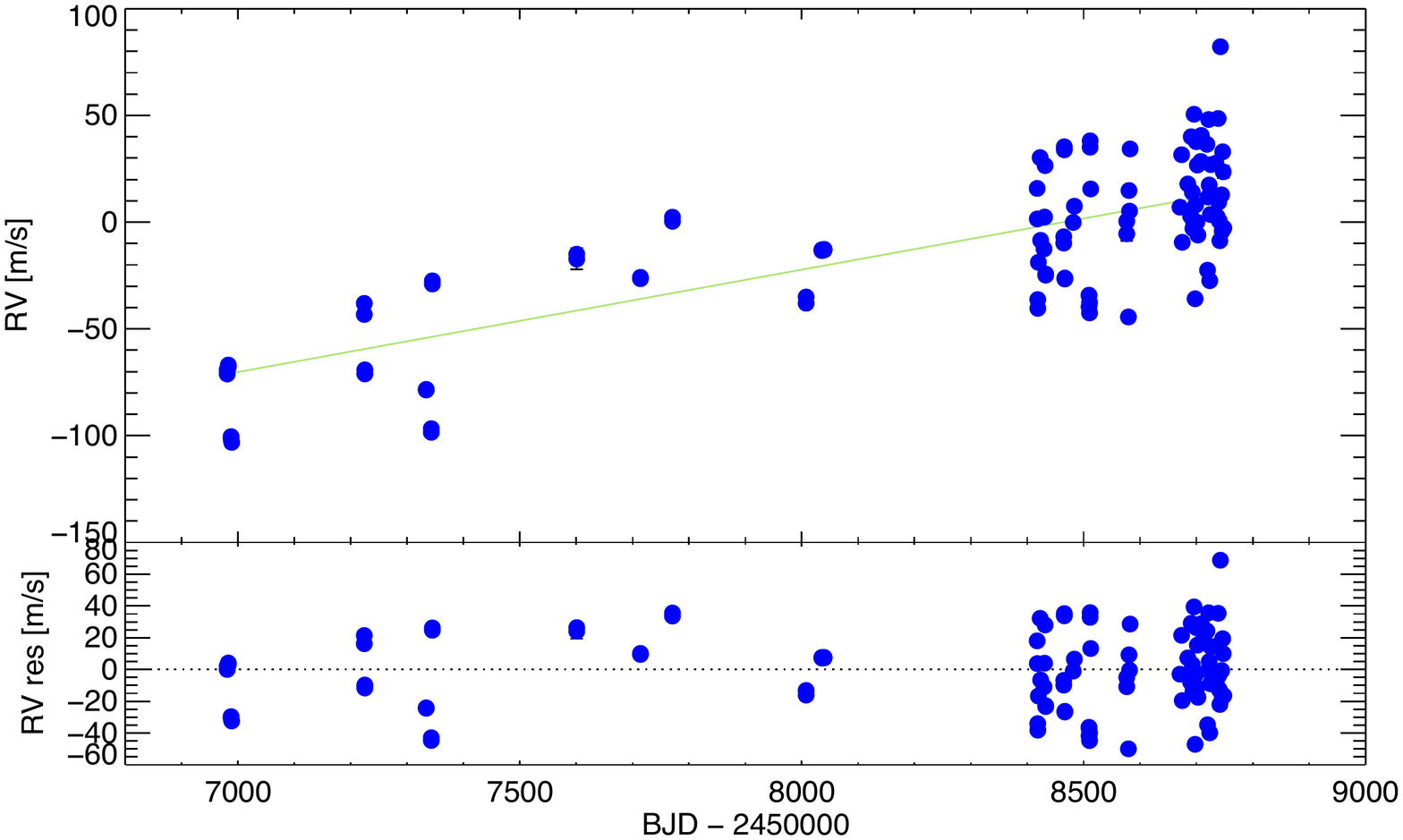}
    \caption{HARPS RV time series of TOI-179. 
    The green line in the upper panel depicts a linear interpolation of the  long-term trend. Lower panel shows the residuals of the model. }
    \label{fig:tsTOI179}
\end{figure}

We measured the RVs by using the Template Enhanced Radial velocity Reanalysis Application (TERRA) pipeline (v1.8) \citep{2012ApJS..200...15A}, which is more effective and less affected by stellar activity features of young and active stars than the cross-correlation function (CCF)-based method \citep{damasso_v830tau}  allowed us to obtain a uniform set of reductions of all the spectra obtained from different observational campaigns. Seven spectra were collected before the fiber upgrade intervention on the HARPS spectrograph occurred in May 2015, which introduced an average RV offset of $\sim 16$ m s$^{-1}$ for G-type slowly-rotating stars, as reported by \citet{locurto2015}. We treated these 7 spectra (\textit{HARPS-pre}) and the remaining 96 (\textit{HARPS-post}) as datasets coming from different instruments, both when extracting the RVs (which includes an additional offset) and modelling the time series. 
The typical values (median) of the RV errors is 1.4 m s$^{-1}$.
Fig.~\ref{fig:tsTOI179} (upper panel) shows the RVs for TOI-179 overimposed by a simple 
linear fit that we used to remove the trend (lower panel) to evaluate the RV dispersion (24.5 m s$^{-1}$), which is likely dominated by the contribution of the stellar activity. For a better visualization, we applied an RV offset to the \textit{HARPS-pre} spectra by using the coefficients of the linear trend.

We also considered the data products of the instrument pipeline \footnote{\url{https://www.eso.org/sci/facilities/lasilla/instruments/harps/tools/archive.html}} to evaluate the examine
 the bisector velocity span (BIS) and  the FWHM of the CCF. 
Finally, we also exploited the HARPS spectra to measure activity indices and to derive the stellar properties (see Sect. \ref{sec:star}, App. \ref{sec:specanalysis}, \ref{sec:chem}, \ref{sec:lithium}, \ref{sec:vsini}, \ref{sec:activity}, and \ref{sec:rot}). The time series are listed in Appendix \ref{a:table}.

\subsection{High-contrast imaging}
\label{sec:ao}

\begin{table*}[!h]
  \caption{List of the main characteristics of the AO observations of TOI-179 used for this work.}\label{t:obsao}
\centering
\begin{tabular}{cccccccccc}
\hline\hline
Date  &  Instr. & Obs. mode & Coronograph & DIMM seeing & $\tau_0$ & wind speed & Field rot. & DIT & Total exp.\\

\hline
2016 Aug 10  & NaCo   &      L'       &      none          &        1.55               &  1.9 ms   & 12.48     &    $16.3^{\circ}$      &     0.2 s          &   888.8 s      \\
2019 Oct 11  & SPHERE & IRDIFS      & N\_ALC\_YJH\_S & 1.18$^{\prime\prime}$ & 2.8 ms & 9.53 m/s &$22.0^{\circ}$ &  96 s &  3072 s \\
2019 Dec 18  & SPHERE & IRDIFS\_EXT & N\_ALC\_YJH\_S & 0.75$^{\prime\prime}$ & 5.2 ms & 9.32 m/s &$22.4^{\circ}$ &  96 s &  3072 s \\
\hline
\end{tabular}
\end{table*}

\subsubsection{SPHERE}
\label{s:spheredata}

We observed TOI-179 with the VLT high-contrast imaging instrument SPHERE \citep{sphere} during the nights of 2019 Oct 11 and 2019 Dec 18 under the open time program 0104.C-0247(B) (P.I. S.~Desidera). The first observation was acquired using the IRDIFS observing mode with IFS \citep{ifs} operating in the Y and J spectral bands (between 0.95 and 1.35~$\mu$m) and with IRDIS \citep{irdis} operating in the H spectral band with the H23 filter pair \citep[wavelength H2=1.593~$\mu$m; wavelength H3=1.667~$\mu$m; ][]{vigan2010}. The second observation was performed exploiting the IRDIFS\_EXT observing mode with IFS operating in Y, J and H spectral bands (between 0.95 and 1.65~$\mu$m) and with IRDIS operating in the K spectral band using the K12 filter pair (wavelength K1=2.110~$\mu$m; wavelength K2=2.251~$\mu$m). The main characteristics of the two observations are listed in Table~\ref{t:obsao}. In both cases, a 3072~s long sequence of coronagraphic images in pupil stabilized mode was performed, including meridian passage to optimize the speckle subtraction through angular differential imaging techniques. Furthermore, short non-coronagraphic images were taken before and after the coronagraphic sequence, to serve as PSF reference to calibrate the contrast curve and to explore the inner $\sim$0.12" for bright companions.
\par
We reduced the data through the SPHERE data center \citep{spheredatacenter} applying the appropriate calibrations following the data reduction and handling \citep[DRH; ][]{2008SPIE.7019E..39P} pipeline. In the IRDIS case, the requested calibrations are the dark and flat-field correction and the definition of the star center. IFS requires, besides to the dark and flat-field corrections, the definition of the position of each spectrum on the detector, the wavelength calibration and the application of the instrumental flat. On the pre-reduced data we then applied speckle subtraction algorithms like TLOCI \citep{2014SPIE.9148E..0UM} and principal components analysis \citep[PCA; ][]{2012ApJ...755L..28S} as implemented in the SPHERE consortium pipeline application called SpeCal \citep{2018A&A...615A..92G} and also described in \citet{2014A&A...572A..85Z} and in \citet{mesa2015} for the IFS case.

\begin{figure}[htbp]
\centering
\includegraphics[width=\columnwidth]{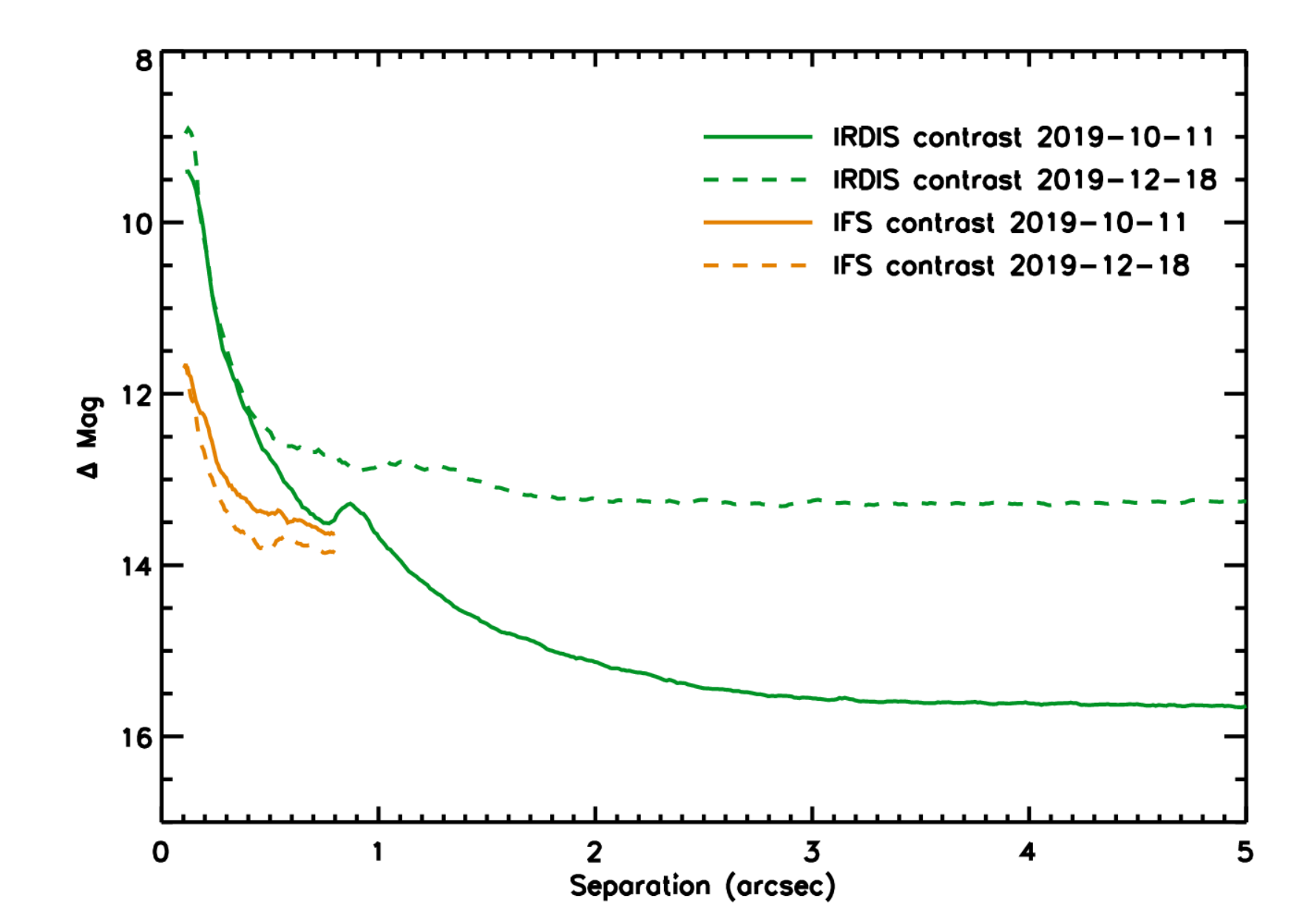}
\caption{\label{fig:magcontrast}  Contrast plot in $\Delta$Mag for TOI-179 for both the SPHERE observing epochs. Green lines are for the IRDIS contrast, orange lines are for the IFS contrast. Solid lines are for the 2019-10-11 epoch, dashed lines are for the
2019-12-18 epoch. } 
\end{figure}

We defined the contrast around the central star for both instruments and for both epochs exploiting the procedure described in \citet{mesa2015} and corrected for the small sample statistic following the method described by \citet{2014ApJ...792...97M}. The final results of this procedure are displayed in Figure~\ref{fig:magcontrast}. 

The non-coronagraphic images obtained with IFS were analyzed using the custom procedure described in \citet{engler2020,shinebin}. This procedure subtracts the static aberrations in the PSF by comparing the non-coronagraphic images obtained before and after the coronagraphic ones, exploiting the field rotation between them. 
A $S/N$ map is then generated using channels at different wavelengths, and position of peaks with $S/N>8$ are then checked. Candidates are kept only if this position is found to be independent of wavelength, since the position corresponding to bright speckles are expected to be wavelength dependent.

\subsubsection{NaCo}
\label{sec:naco}

The star was observed with NaCo on 2016 Aug 10 (see Table~\ref{t:obsao}) in the context of the open time program
097.C-0972(A) (P.I. J. Girard). The observation was performed in L' band
without coronograph and in pupil stabilized mode with a total rotation of the FOV of $16.3^{\circ}$. The dataset was composed by 44 data-cubes, each with 101 frames. The exposure time for each frame was 0.2~s for a total observing time of 888.8~s (less than 15 minutes). The star PSF was in different positions across the detector during the observations allowing to effectively handle the bad pixels. \par
We created a master flat by making a median on the whole dataset and we subtracted the resulting image from each frame of the raw data. We then performed on the reduced dataset a simple
Angular Differential Imaging (ADI) procedure to reduce the stellar noise.

\begin{figure}[htbp]
\centering
\includegraphics[width=\columnwidth]{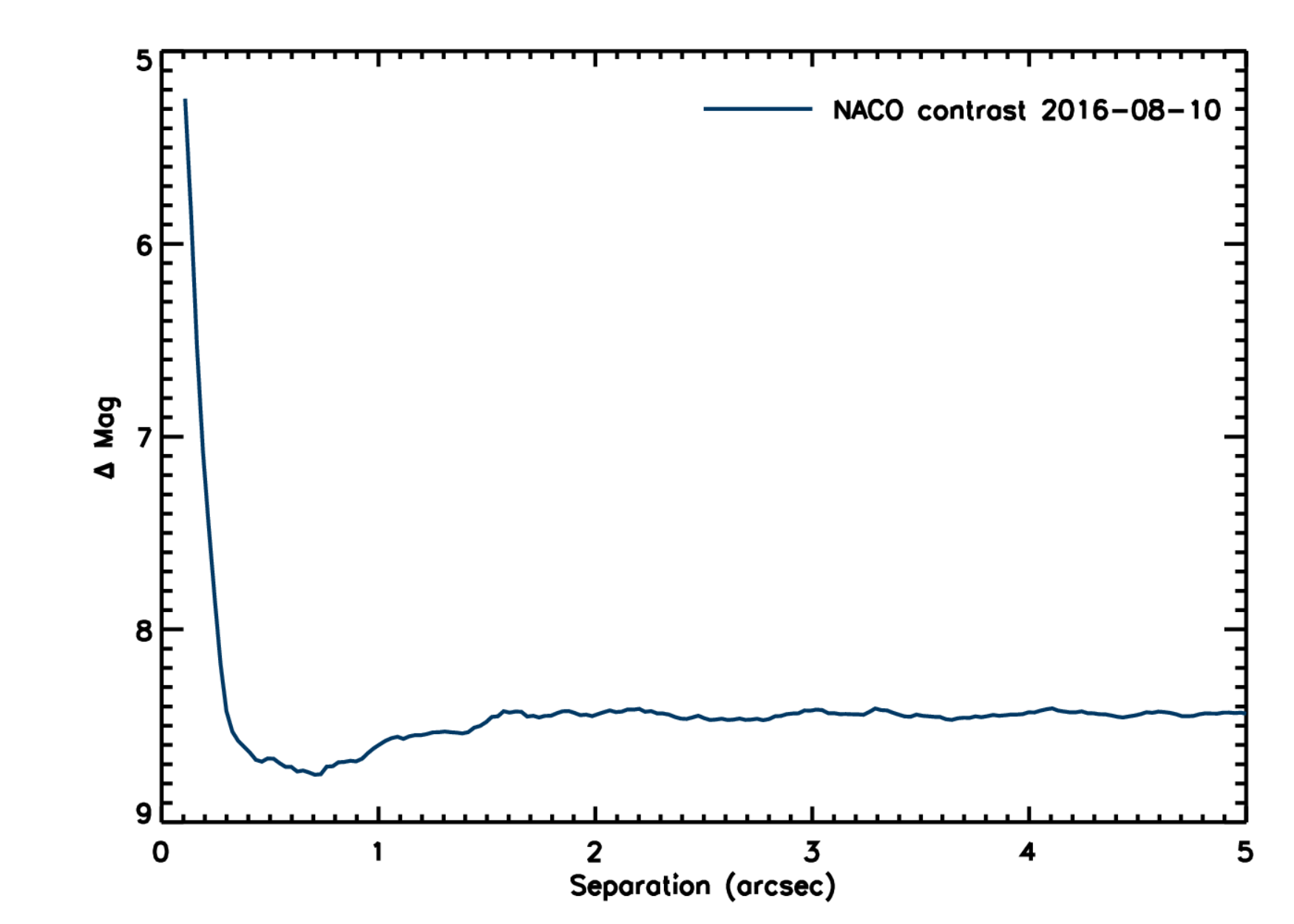}
\caption{\label{fig:magcontrastnaco} Contrast plot in $\Delta$Mag for TOI-179 obtained using the NACO data.  }
\end{figure}

The procedure to infer contrast limits was also exploited on the NaCo data. The results are displayed in Figure~\ref{fig:magcontrastnaco}.

%

\subsection{Chandra observations} 
\label{sec:chandra}

Chandra observed TOI-179 in two visits (OBsID 13481 and 14448, at epochs 2012 Jun 23 and  2012 Jun 24, respectively) for 22 and 26 ks, respectively.
We downloaded the latest calibrated datasets and event lists and extracted the spectra and light curves
of TOI-179 and its wide companion \object{CD-56 583} (see Sect.~\ref{sec:bin}) with CIAO 4.14 and {\it dmextract}. We defined two circular regions of about
25$\arcsec$ of radius to extract the events related to each source and background. 
For CD-56 583 the source extraction region includes both components.
The task {\it dmextract} accumulates the spectra of source and background, calculates the response files and 
groups the spectra to a minimum of 20 counts per bin in a way that they are suitable for the best fit 
analysis. We used XSPEC v12.11.1b to analyze the grouped spectra within which we defined a model 
of coronal emission composed by a thermal optically thin plasma (APEC) absorbed by a global equivalent 
H column (TBAbs). We set a fixed low absorption N$_H$ value of $10^{17}$ cm$^{-2}$ and metals abundances 
pattern ($Z=0.2Z_\odot$) and let free to vary kT and spectrum normalization.
The spectra from both observations were fitted simultaneously. 
These data are used in App.~\ref{sec:activity} and \ref{sec:bin} to
infer the X-ray luminosity of the components of the system.

\section{Summary of stellar parameters}
\label{sec:star}

TOI-179 was previously recognized as an active star \citep{jenkins2008,jenkins2011} but with limited
studies in the literature so far. It was included in the RV search for planets around young stars carried out by \citet{grandjean2020}. It was also in the sample
of SPHERE SHINE GTO survey as a low-priority target \citep{desidera2021} but left unobserved at the end of the survey.

We performed our own determination of stellar parameters, exploiting the HARPS spectra, the TESS photometric time series, and the Chandra observations described in Sect. \ref{sec:obs}, coupled with additional data and results from the literature.
These determinations are discussed in detail in Appendix \ref{app:star} and the adopted parameters are summarized in Table \ref{t:star_param}.

In brief, the age indicators of TOI-179 consistently support an age older than the Pleiades and younger than the Hyades. After comparison with the 300 Myr old Group X association \citep{messina2022}, we infer an age of 400 Myr with limits between 300 to 500 Myr. The star is an early K dwarf (App. \ref{sec:specanalysis})  with roughly solar chemical composition,  as resulting from the
analysis of several elements performed in App. \ref{sec:chem}. The stellar mass and radius result 0.863$\pm$0.020~M$_\odot$ and  0.767$\pm$0.024~R$_\odot$, respectively.
An edge-on orientation is derived for the star from the observd rotation period, projected rotation velocity, and stellar radius, suggesting alignment with respect to the orbit of the transiting planet TOI-179b. Finally, TOI-179 forms a triple system with the close pair of late K dwarfs CD-56 593A and B (relative separation $\sim$ 1"), at a projected separation of 87.5" from TOI-179 corresponding to 3400 au at the distance of the system (38 pc), as
described in App.~\ref{sec:bin}. The physical association is confirmed by Gaia astrometry.
The system does not result a member of groups or associations, as found in  the analysis performed in App.~\ref{sec:kin}.

\begin{table}[!htb]
   \caption[]{Stellar properties of TOI-179}
     \label{t:star_param}
     \small
     \centering
       \begin{tabular}{lcc}
         \hline
         \noalign{\smallskip}
         Parameter   &  TOI-179 & Ref  \\
         \noalign{\smallskip}
         \hline
         \noalign{\smallskip}
$\alpha$ (J2000)         &  02 57 02.95        & Gaia EDR3    \\
$\delta$ (J2000)         & -56 11 31.51        & Gaia EDR3  \\
$\mu_{\alpha}$ (mas/yr)  &  -36.661$\pm$0.015  & Gaia EDR3  \\
$\mu_{\delta}$ (mas/yr)  &   50.558$\pm$0.015  & Gaia EDR3  \\
RV     (km/s)            &   -0.48$\pm$0.26    & Gaia DR2   \\
RV     (km/s)            &   -0.20$\pm$1.70    & \citet{jenkins2011}    \\
$\pi$  (mas)             &  25.8847$\pm$0.0128 & Gaia EDR3  \\
$U$   (km/s)             &  -3.562$\pm$0.011   & \citet{smart2021}  \\
$V$   (km/s)             &   8.924$\pm$0.151   & \citet{smart2021}  \\
$W$   (km/s)             &  -6.223$\pm$0.203   & \citet{smart2021}  \\
\noalign{\medskip}
V (mag)                  &   8.99             & Hipparcos  \\
B-V (mag)                &   0.876$\pm$0.005  & Hipparcos  \\
V-I (mag)                &   0.92$\pm$0.01    & Hipparcos  \\
G (mag)                  & 8.7312$\pm$0.0007  & Gaia DR2  \\
BP-RP (mag)              &  1.073             & Gaia DR2  \\
J$_{\rm 2MASS}$ (mag)    &   7.428$\pm$0.018  & 2MASS  \\
H$_{\rm 2MASS}$ (mag)    &   7.029$\pm$0.015  & 2MASS  \\
K$_{\rm 2MASS}$ (mag)    &   6.883$\pm$0.020  & 2MASS  \\
T (mag)                  &   8.180$\pm$0.006  & TIC  \\
\noalign{\medskip}
Spectral Type            & K2V                &  \citet{gray2006} \\
T$_{\rm eff}$ (K)        &  5172$\pm$60       & This paper (spec) (App. \ref{sec:specanalysis}) \\
T$_{\rm eff}$ (K)        &  5118$\pm$60       & This paper (phot) (App. \ref{sec:specanalysis}) \\
T$_{\rm eff}$ (K)        &  5145$\pm$50       & This paper (adop.) (App. \ref{sec:specanalysis})\\
$\log g$                 &  4.54$\pm$0.09     & This paper  (App. \ref{sec:specanalysis}) \\ 
${\rm [Fe/H]}$ (dex)     &  0.00$\pm$0.08     & This paper (App. \ref{sec:specanalysis})  \\ 
\noalign{\medskip}
$S_{\rm MW}$             &   0.606$^{+0.067}_{-0.086} $    & This paper  (App. \ref{sec:activity}) \\
$\log R^{'}_{\rm HK}$    &     -4.35$^{+0.05}_{-0.08} $ &  This paper  (App.  \ref{sec:activity})  \\  
$ v \sin i $ (km/s)      &    4.5$\pm$0.5   & This paper (App. \ref{sec:vsini}) \\  
${\rm P_{\rm rot}}$ (d)  &   8.73$\pm$0.07  &   This paper  (App. \ref{sec:rot})   \\
$\log L_{\rm X}$             &  28.51   & This paper (App.  \ref{sec:activity})  \\
$\log L_{\rm X}/L_{\rm bol}$     &  -4.65  & This paper ( App.  \ref{sec:activity})  \\
$EW_{\rm Li}$ (m\AA)        &  39.3$\pm$4.5 &  This paper  (App. \ref{sec:lithium})   \\
A(Li)                          &  1.55$\pm$0.08 &  This paper (App. \ref{sec:lithium})  \\
\noalign{\medskip}
Mass (M$_{\odot}$)       &   0.863$\pm$0.020 & This paper  (App. \ref{a:massradius})  \\
Radius (R$_{\odot}$)     &   0.767$\pm$0.024  & This paper (App. \ref{a:massradius})  \\
Luminosity (L$_{\odot}$) &    0.372$\pm$0.012 & This paper (App. \ref{a:massradius} )  \\
Age  (Myr)               & 400$\pm$100 & This paper (App. \ref{a:age})   \\
         \noalign{\smallskip}
         \hline
      \end{tabular}
\end{table}


\section{Confirmation of a low-mass transiting planet around TOI-179} \label{sec:toi179b}

\subsection{TESS light curve analysis
\label{sec:phot_toi179}}

\begin{figure*}
    \centering
    \includegraphics[bb=17 250 570 710, width=0.95\textwidth]{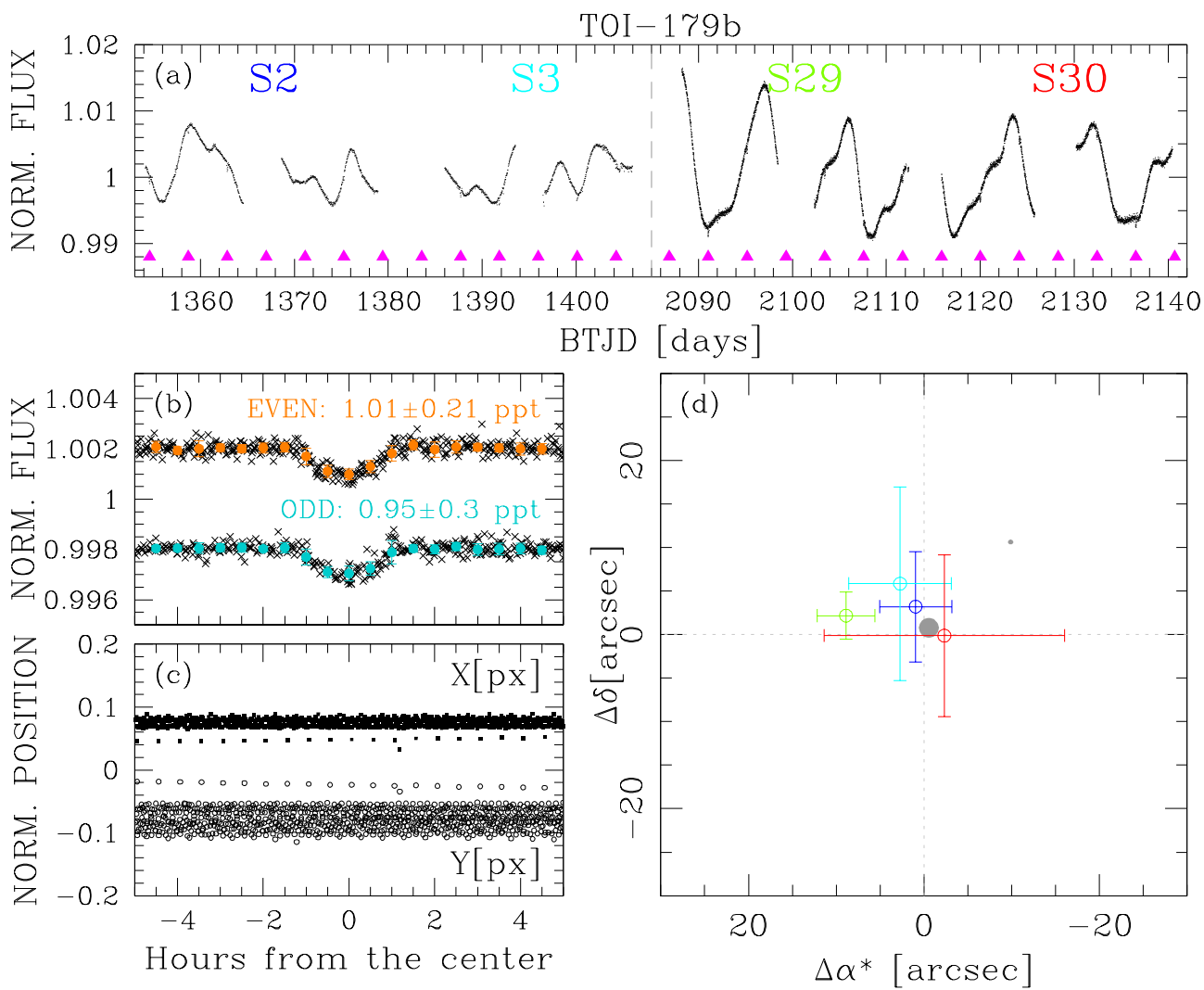}
    \caption{Overview on the vetting tests performed for TOI-179b by using the {\it TESS} data. Panel {\it (a)}  shows the flux-normalized light curve of TOI-179 obtained by using the FFIs collected during Sectors 2, 3, 29, and 30: magenta triangles indicate the transit events. Panel {\it (b)} shows the comparison between phased odd and even transits: orange/cyan points are the binned phased light curves, with bins of 0.5 hours. Within the errors, odd and even transits have the same depths. Panel{\it (c)} shows the phased X-/Y-positions: there is no correlation between the positions of the star in each image and the transit signals. Panel {\it (d)} illustrates the in-/out-of transit difference centroid analysis: within $<3\sigma$ the centroids correspond to the position of the star; each centroid is color-coded as the name of the sectors in panel {\it (a)}.}
    \label{fig:toi179vet}
\end{figure*}

The candidate exoplanet TOI-179b was detected for the first time by the  Quick-Look Pipeline (QLP, \citealt{2020RNAAS...4..204H}) by using data from Sectors 2 and 3. By using the light curves extracted from the FFIs as described in Sect.~\ref{sec:tess}, we performed a series of vetting tests based on the {\it TESS} data to confirm this candidate exoplanet. A complete description of these tests are reported in \citet{2020MNRAS.498.5972N}: first, we modelled and removed the variability of the star from the light curve by interpolating it with a 5th-order spline defined on a series of knots spaced 6.5 hours; during this procedure, we also removed all the points with the photometric flag \texttt{DQUALITY>0}. We detected the transit signals of the candidate exoplanets by extracting the Transit Least Squares (TLS) periodogram (\citealt{2019A&A...623A..39H}) of the flattened light curve, searching for transits between 0.5 days and 150.0 days. In this way we confirmed the periodic transit signal at $P\sim 4.1374$~d reported by the QLP report. An overview of the transit times for TOI-179b is reported in panel (a) of Fig.~\ref{fig:toi179vet}. The same figure shows that odd and even transits have the same depth within the errors (panel (b)) and that there is no correlation between the transit events and the X/Y-position of the star calculated by fitting to it the PSF in each image (panel (c)). Even if the star is isolated, we performed the analysis of in-/out-of transit difference centroid (panel (d)): within the errors ($<3\sigma$), the mean centroids calculated for the different sectors coincide with the position of the star.

\subsection{Frequency content analysis of the radial velocities} \label{sec:glsrv}

\begin{figure}
    \centering
    \includegraphics[width=0.50\textwidth]{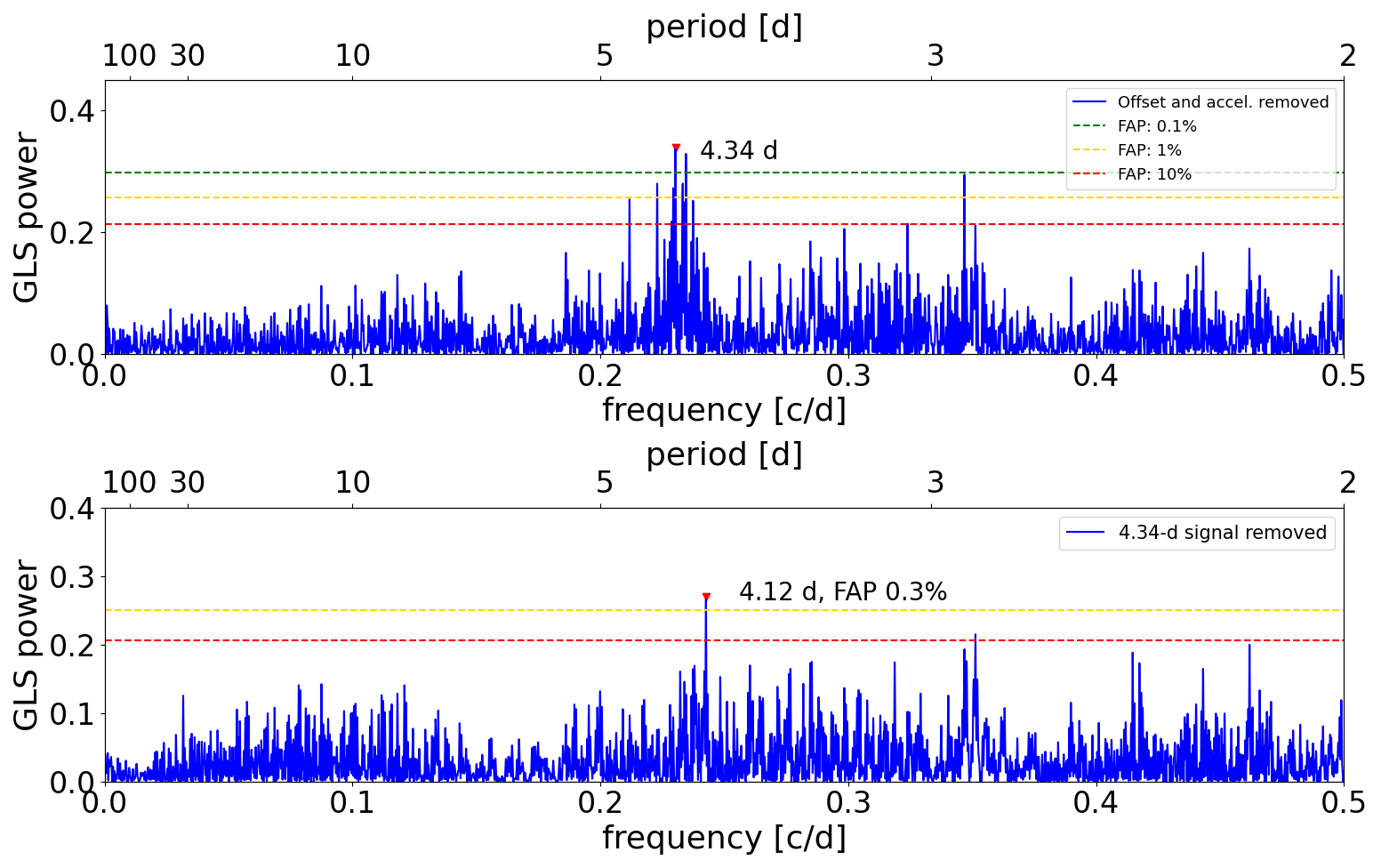}
    \caption{\textsc{GLS} periodograms of the RVs: after removing the long-term acceleration from the original data (upper panel); for the pre-whitened data, after removing the best-fit sinusoid calculated by \textsc{GLS} (lower panel). Both frequencies (bottom axis) and periods (top axis) are shown on the x-axis. Bootstrap-derived FAPs are indicated by horizontal dashed lines.}
    \label{fig:glsrv}
\end{figure}

Before performing a more complex analysis, we investigated the frequency content in the RV time series by calculating the generalised Lomb-Scargle (\textsc{GLS}, \citealt{zechmeister2009}) periodogram. We were particularly interested in inspecting the presence of periodic signals ascribable to stellar rotation and planetary orbit. Before running the \textsc{GLS} algorithm, we removed an offset and a linear long-term trend form the data, adopting the best-fit values for $\gamma_{\rm HARPS-pre}$, $\gamma_{\rm HARPS-post}$, and $\dot{\gamma}$ listed in Table \ref{t:toi-179_pl_param}. The results are shown in Fig. \ref{fig:glsrv} (upper panel). The periodogram shows the main peak at $P=4.34$ days, with a significantly low false alarm probability (FAP), that we derived through a bootstrap analysis. This period corresponds to the first harmonic of the stellar rotation period. Another significant peak (FAP$\sim1\%$) is observed at a higher frequency, corresponding to the second harmonic of the stellar rotation period. 
Subsequently, we pre-whitened the data by removing the best-fit sinusoid model obtained with \textsc{GLS}, and we calculated the periodogram of the residuals, that is shown in the lower panel of Fig.~\ref{fig:glsrv}. The main peak is located very close to the orbital period of the transiting companion to TOI-179, as measured by \textit{TESS}, and it appears significant (FAP=0.3$\%$). 
We conclude that, from the frequency content analysis we obtained two results which are useful for the more complex analysis described in Section \ref{sec:rv_tess_fit_toi179}: signals related to stellar activity dominate the RV time series, and the signal due to the transiting companion is already detected after pre-whitening the data. 

\subsection{Joint modelling of radial velocities and photometric transits } 
\label{sec:rv_tess_fit_toi179}

We calculated and derived the main planetary parameters of TOI-179\,b by performing a fit of the combined RV+TESS short-cadence photometry (sectors 2--3 and 29--30) dataset. Instead of the full TESS data set, we used that part of the flattened light curve containing only the transit signal, and a long out-of-transit baseline ($\sim$5 hr before and after the transit) to guarantee a proper modelling of the photometric data. We explored the full (hyper-)parameter space using the publicly available Monte Carlo (MC) nested sampler and Bayesian inference tool \textsc{MultiNest v3.10} (e.g. \citealt{Feroz2019}), through the \textsc{pyMultiNest} wrapper \citep{Buchner2014}. Our MC set-up included 500 live points, and we adopted a sampling efficiency of 0.3. We used the code \textsc{batman} \citep{Kreidberg2015} for modelling the photometric transits. Concerning the RV data, we modelled the time variability dominated by stellar activity modulations by means of a Gaussian process (GP) regression, and adopting a quasi-periodic (QP) kernel. We used the publicly available \textsc{python} module \textsc{george} v0.2.1 \citep{2015ITPAM..38..252A} to perform the GP regression within the MultiNest framework. The elements of the QP covariance matrix (e.g. \citealt{haywood14}) implemented in our model is defined as follows:

\begin{eqnarray} 
\label{eq:eqgpqpkernel}
k_{QP}(t, t^{\prime}) = h^2\cdot\exp\Bigg[-\frac{(t-t^{\prime})^2}{2\lambda^2} - \frac{\sin^{2}\bigg(\pi\big(t-t^{\prime})/\theta\bigg)}{2w^2}\Bigg] + \nonumber \\
+\, (\sigma^{2}_{\rm RV}(t)+\sigma^{2}_{\rm jit,\,RV})\cdot\delta_{t, t^{\prime}}
\end{eqnarray}

Here, $t$ and $t^{\prime}$ represent two different epochs of observations, $\sigma_{\rm RV}$ is the radial velocity uncertainty, and $\delta_{t, t^{\prime}}$ is the Kronecker delta. Our analysis takes into account other sources of uncorrelated noise -- instrumental and/or astrophysical -- by including a constant jitter term $\sigma_{\rm jit,\, RV}$ which is added in quadrature to the formal uncertainties $\sigma_{\rm RV}$. The GP hyper-parameters are $h$, which denotes the scale amplitude of the correlated signal; $\theta$, which represents the periodic time-scale of the modelled signal, and corresponds to the stellar rotation period; $w$, which describes the "weight" of the rotation period harmonic content within a complete stellar rotation (i.e. a low value of $w$ indicates that the periodic variations contain a significant contribution from the harmonics of the rotation periods); and $\lambda$, which represents the decay timescale of the correlations, and is related to the temporal evolution of the magnetically active regions responsible for the correlated signal observed in the RVs. 
When fitting the RVs, we introduced two offsets ($\gamma_{pre}$ and $\gamma_{post}$) and white noise jitters ($\sigma_{pre}$ and $\sigma_{post}$) for the pre- and post-fiber upgrade datasets.
Concerning the modelling of the TESS transit light curve, we adopted a limb darkening (LD) quadratic law, with uniform priors on the LD coefficients LD$_{\rm c1}$ and LD$_{\rm c2}$. We fitted constant jitter terms $\sigma_{\rm jit,\, TESS}$ in quadrature to the formal photometric uncertainties, one jitter for data collected in sectors 2-3, and one for those from sectors 29-30, to take into account variations in TESS performance over two years of operation. 
We modelled the orbit of planet b with a Keplerian, adopting the parametrization $\sqrt{e_{b}}\cos\omega_{b,\,\star}$ and $\sqrt{e_{b}}\sin\omega_{b,\,\star}$ instead of using $e_{b}$ and $\omega_{b,\,\star}$ as free parameters. We included an acceleration $\dot{\gamma}$ to model the long-term trend that dominates the RV time series, and also tested a parabolic long-term trend by including the additional free parameter $\ddot{\gamma}$. 

\begin{figure}
    \centering
    \includegraphics[width=0.45\textwidth]{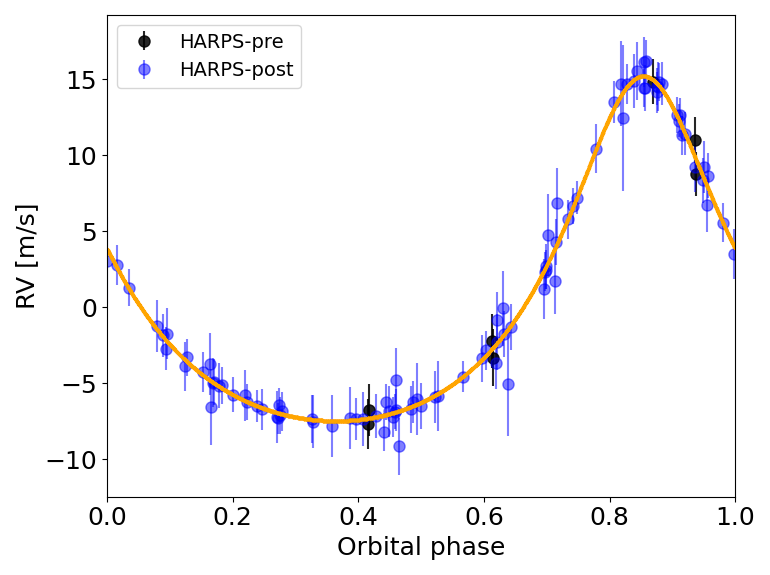}
    \caption{Spectroscopic orbit of TOI-179 due to planet b. The best-fit model is indicated by an orange line.}
    \label{fig:spectorbi}
\end{figure}

\begin{figure}[!h]
    \centering
    \includegraphics[width=0.5\textwidth]{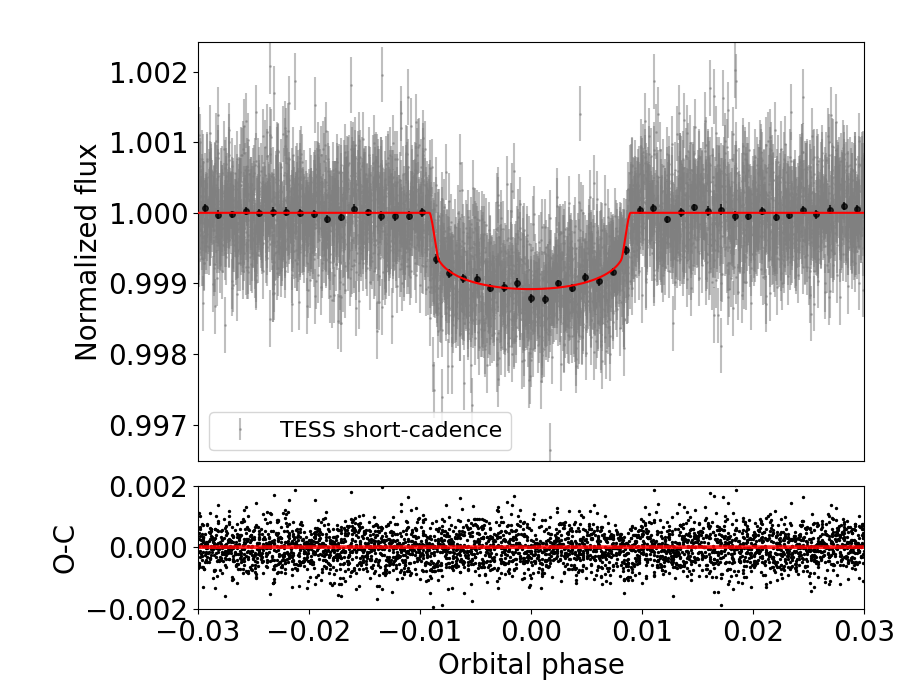}
    \caption{Best-fit transit signal of TOI-179~b based of TESS short-cadence photometry (Tab. \ref{t:toi-179_pl_param}). Model residuals O-C are shown in the bottom panel.}
    \label{fig:tess_transit_scad}
\end{figure}

Priors and best-fit values for all the free (hyper-)parameters used in our model are summarized in Tab.~\ref{t:toi-179_pl_param} while cornerplots of the resulting distributions and correlations of each parameter are shown in App.~\ref{appendix:cornerplot}.
Our results show that the model including a linear long-term trend in the RVs is statistically much more favoured over a model with a quadratic trend ($\ln\mathcal{Z}_{linear}-\ln\mathcal{Z}_{quadr.}=+10.2$, where $\mathcal{Z}$ represents the Bayesian evidence), excluding the presence of a 
significant curvature in the RVs. The RV signature of the transiting planet is detected with a 3.3$\sigma$ confidence ($K=11.3^{+3.3}_{-3.6}$ m/s), corresponding to a mass $m_b=24.1^{+7.1}_{-7.7}$ $M_{\oplus}$, indicating a rather compact structure for the planet.
The planetary orbit is eccentric with a 3.8$\sigma$ significance ($e_b=0.34^{+0.07} _{-0.09}$). These properties are further discussed in 
Sect.~ \ref{sec:discussion}.
Fig. \ref{fig:spectorbi} and \ref{fig:tess_transit_scad} show the corresponding best-fit spectroscopic orbit of TOI-179 due to planet b, the and transit light curve of TOI-179\,b.

\begin{figure*}[!h]
    \centering
    \includegraphics[width=0.8\textwidth]{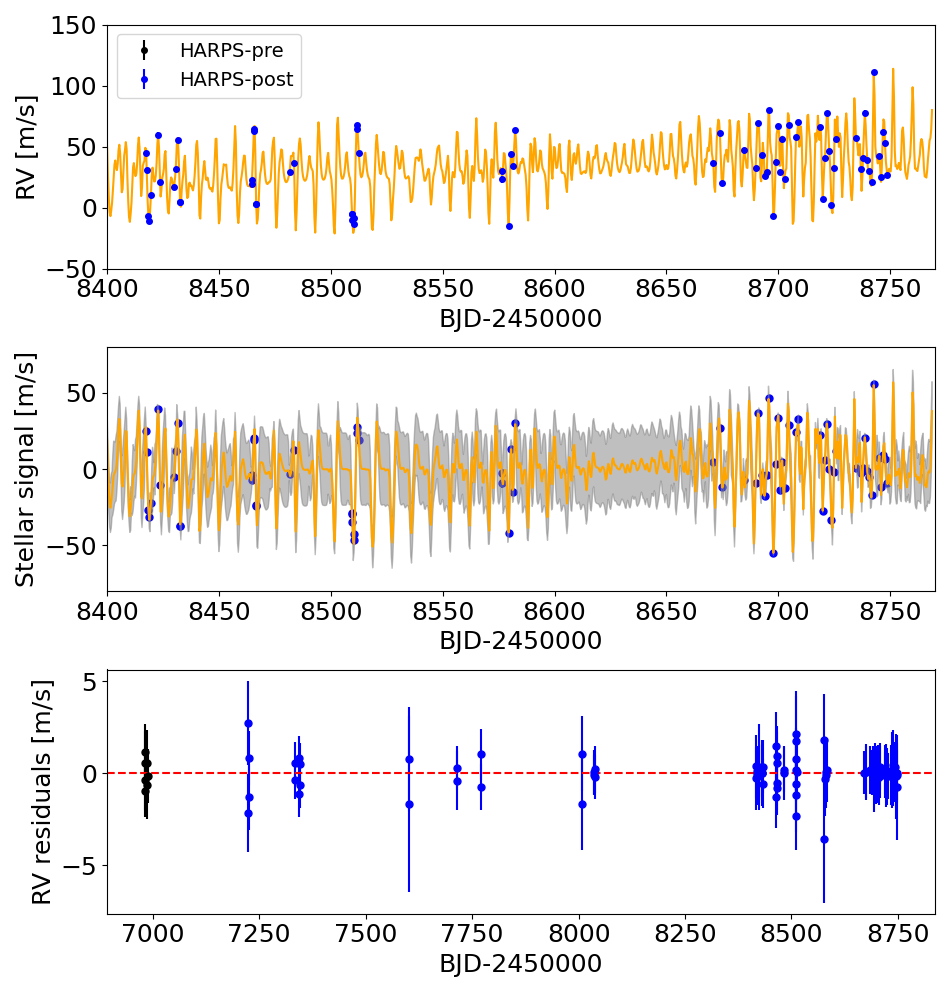}
    \caption{Best-fit results of the joint RV+TESS short-cadence light curve (Tab. \ref{t:toi-179_pl_param}). \textit{First panel:} Original RVs with offsets removed (only HARPS-post RVs are shown for clarity). \textit{Second panel:} Quasi-periodic stellar activity term (only HARPS-post RVs are shown for clarity). \textit{Last panel:} RV residuals.
    The RV have been extracted with the pipeline TERRA, and error bars include the uncorrelated jitter terms added in quadrature to the formal uncertainties. The orange lines indicate the best-fit models: Keplerian+linear trend+stellar activity (upper panel); stellar activity (second panel). The grey shaded area in the second panel marks the $\pm1\sigma$ uncertainty of the quasi-periodic activity signal.}
    \label{fig:rv_orbit_scad}
\end{figure*}

We note that the results for the GP part of the RV model can be considered realistic and trustworthy. The stellar rotation period $\theta$ is precisely recovered even using a quite broad uninformative prior, and the short decay timescale $\lambda$, equivalent to about three rotation cycles (see Fig.~\ref{fig:cornerplots}), is supported by the result we got after applying a GP quasi-periodic regression to the TESS photometry of sectors 2-3 and 29-30, modelled separately and with transit signals removed. We found $\lambda\sim10$ days in both cases, showing that the evolutionary timescale of the active regions is indeed of the order of a few stellar rotational periods. It must be taken into account that the TESS and RVs measurements are not contemporaneous, hence the stellar activity behavior cannot be compared directly in the two dataset (for that reason we could not use the photometry to directly constrain the activity term in the RV model). Moreover, the two datasets have very different sampling, therefore we cannot expect to find the same values of $\lambda$, but only to verify that the order of magnitude is similar to support the result for the RVs, as it happens in this case.

\begin{table*}[htbp]
   \caption{Best-fit values and priors of the model parameters for the combined RV+TESS photometry fit of TOI-179 discussed in Sect. \ref{sec:rv_tess_fit_toi179}.}
    \label{t:toi-179_pl_param}
     \footnotesize
     \begin{center}
       \begin{tabular}{llll}
         \hline
         \noalign{\smallskip}
         Parameter   & Prior\tablefootmark{a} & \multicolumn{2}{c}{Best-fit value\tablefootmark{b}}\\
         \noalign{\smallskip}
         \hline
         \noalign{\smallskip}
         & & RV linear trend (adopted) & RV parabolic trend  \\
         \noalign{\smallskip}
         \hline
         \noalign{\smallskip} 
\textbf{Fitted} & & & \\
         \noalign{\smallskip}
\textit{RV stellar activity term:} & & & \\
         \noalign{\smallskip}
            $h$ [\ms] & $\mathcal{U}$(0,50)  & $23.8^{\rm +2.4}_{\rm -2.2}$ & $24.0^{\rm +2.6}_{\rm -2.1}$ \\
            \noalign{\smallskip}
            $\lambda$ [d] & $\mathcal{U}$(0,1000) & 24.0$^{\rm +6.7}_{\rm -6.2}$ & 24.1$^{\rm +6.8}_{\rm -6.2}$ \\
            \noalign{\smallskip}
            $w$ & $\mathcal{U}$(0,1) & $0.16\pm0.03$ & $0.16\pm0.03$ \\ 
            \noalign{\smallskip}
            $\theta$ [d] & $\mathcal{U}$(3,10) & $8.72\pm0.04$ & $8.71\pm0.04$ \\
            \noalign{\smallskip}
\textit{Planet-related parameters:} & & & \\
         \noalign{\smallskip}  
          $K_b$ [\ms] & $\mathcal{U}$(0,50) & 11.3$^{\rm +3.3}_{\rm -3.6}$ & 11.3$^{\rm +3.4}_{\rm -3.6}$ \\
          \noalign{\smallskip} 
            orbital period, $P_b$ [d] & $\mathcal{U}$(4,4.2)  & $4.1374354^{+0.0000036}_{-0.0000037}$  & $4.1374355\pm3.6\cdot10^{-6}$ \\
            \noalign{\smallskip}
            T$_{conj,\,b}$ [BJD-2450000] & $\mathcal{U}$(9111.68,9111.78)  & 9111.73946$^{\rm +0.00068}_{\rm -0.00057}$ & 9111.73948$^{\rm +0.00072}_{\rm -0.00058}$\\
            \noalign{\smallskip}
            $\sqrt{e_{b}}\cos\omega_{b,\,\star}$  & $\mathcal{U}$(-1,1)  &  0.56$^{\rm +0.05}_{\rm -0.08}$ & 0.56$^{\rm +0.06}_{\rm -0.08}$\\
            \noalign{\smallskip} 
            $\sqrt{e_{b}}\sin\omega_{b,\,\star}$  & $\mathcal{U}$(-1,1)  & -0.025$^{\rm +0.151}_{\rm -0.183}$ & -0.033$^{\rm +0.158}_{\rm -0.179}$ \\
            \noalign{\smallskip}             
            acceleration, $\dot{\gamma}$ [$\ms d^{-1}$] & $\mathcal{U}$(0,0.1)  &  $0.048\pm0.008$ &  $0.047\pm0.011$ \\
            \noalign{\smallskip} 
             curvature, $\ddot{\gamma}$ [$\ms d^{-2}$] & $\mathcal{U}$(-1,1)  &  - &  $0.4\pm2.2\cdot10^{-5}$ \\
            \noalign{\smallskip} 
            $R_{\rm b}/R_{\star}$ & $\mathcal{U}$(0.02,0.05) & 0.0309$^{\rm +0.0016}_{\rm -0.0012}$ & 0.0310$^{\rm +0.0016}_{\rm -0.0012}$ \\
            \noalign{\smallskip}
            inclination, $i_b$ [deg] & $\mathcal{U}$(80,90)  &  87.6$^{\rm +1.4}_{-1.0}$ & 87.5$^{\rm +1.5}_{-0.9}$ \\
         \noalign{\medskip}
\textit{RV-related parameters:} & & & \\
         \noalign{\smallskip}          
            $\sigma_{\rm jit,\: HARPS-pre}$ [\ms] & $\mathcal{U}$(0,20) & 1.2$_{\rm -0.8}^{\rm +1.6}$ & 1.2$_{\rm -0.8}^{\rm +1.6}$ \\ 
            \noalign{\smallskip}
            $\gamma_{\rm HARPS-pre}$ [\ms] & $\mathcal{U}$(-100,+100) & 26.2$_{\rm -14.3}^{\rm +14.5}$ & 21.3$_{\rm -27.2}^{\rm +27.7}$ \\
            \noalign{\smallskip}
            $\sigma_{\rm jit,\: HARPS-post}$ [\ms] & $\mathcal{U}$(0,20) & 0.4$_{\rm -0.3}^{\rm +0.5}$ & 0.4$_{\rm -0.3}^{\rm +0.5}$ \\ 
            \noalign{\smallskip}
            $\gamma_{\rm HARPS-post}$ [\ms] & $\mathcal{U}$(-100,+100) & -29.3$_{\rm -5.4}^{\rm +5.6}$  & -$30.3_{\rm -8.3}^{\rm +7.8}$\\
            \noalign{\smallskip}
\textit{Light curve-related parameters:} & & & \\  
         \noalign{\smallskip}
         $\sigma_{jit, \,TESS\,sect.\,2-3}$ &  $\mathcal{U}$(0,0.001) &  $0.000177\pm0.000009$ &  $0.000177\pm 0.000009$ \\
         \noalign{\smallskip}
         $\sigma_{jit, \,TESS\,sect.\,29-30}$ &  $\mathcal{U}$(0,0.001) &  $0.000185\pm0.000009$ &  $0.000185\pm0.000009$ \\
         \noalign{\smallskip}
         LD$_{\rm c1}$ &  $\mathcal{U}$(0,1) & $0.35^{+0.17}_{-0.20}$ & $0.35^{+0.17}_{-0.20}$ \\
         \noalign{\smallskip}
         LD$_{\rm c2}$ &  $\mathcal{U}$(0,1) & $0.34^{+0.31}_{-0.24}$ & $0.33^{+0.32}_{-0.23}$ \\
         \noalign{\smallskip}
\textbf{Derived} & & & \\
         \noalign{\smallskip}
         $a_{b}$ [au] & --  &  $0.0480\pm0.0004$ &  $0.0480\pm0.0004$ \\
         \noalign{\smallskip}
         $e_{b}$  & --  &  0.34$_{\rm -0.09}^{\rm +0.07}$ & 0.34 $_{\rm -0.09}^{\rm +0.08}$\\
         \noalign{\smallskip}
         $\omega_{b,\,\star}$ [rad] & --  &  -0.044 $_{\rm -0.311}^{\rm +0.284}$ & -0.06 $\pm$0.03 \\
         \noalign{\smallskip} 
         $a_b/R_{\star}$\tablefootmark{c} & --  &  14.1$_{\rm -1.5}^{\rm +1.0}$ &  14.0$_{\rm -1.4}^{\rm +1.1}$ \\
         \noalign{\smallskip}
         impact param., $b$ & --  &  0.50$_{\rm -0.26}^{\rm +0.12}$ &  0.51$_{\rm -0.28}^{\rm +0.11}$ \\
         \noalign{\smallskip}
         transit duration, $T_{1,4}$ [d] & --  &  0.0860$_{\rm -0.0053}^{\rm +0.0051}$ &  0.0861$_{\rm -0.0051}^{\rm +0.0056}$ \\
         \noalign{\smallskip}
         radius, $R_b$ [R$_{\rm \oplus}$]  &  -- & 2.60$_{\rm +0.15}^{\rm -0.13}$ & 2.60$_{\rm -0.13}^{\rm +0.15}$ \\
         \noalign{\smallskip}
         mass, $m_b$ [M$_{\rm \oplus}$] & -- & 24.1$^{+7.1}_{-7.7}$ & 24.0$^{+7.3}_{-7.6}$ \\
         \noalign{\smallskip}
         mean density, $\rho_b$ [g cm$^{\rm -3}$] & -- & 7.5$^{+2.6}_{-2.4} $ & 7.4 $^{+2.7}_{-2.5} $ \\
         \noalign{\smallskip}
         surface gravity, $g_{\rm b }$ [m s$^{-2}$] & -- & $35\pm11$ & $35\pm11$ \\
         \noalign{\smallskip}
         \hline
         \noalign{\smallskip}
         Log. Bayesian evidence, $\ln\mathcal{Z}$ &       & 13971.5 & 13961.3 \\ 
         \noalign{\smallskip}
         \hline
      \end{tabular}
      \tablefoot{
      \tablefoottext{a}{$\mathcal{U}$ denotes a prior drawn from an uninformative distributions.}
      \tablefoottext{b}{Parameter uncertainties are given as the $16^{\rm th}$ and $84^{\rm th}$ percentiles of the posterior distributions. }
      \tablefoottext{c}{We used the stellar density $\rho_{*}$ [$\rho_{\odot}$] as free parameter with prior $\mathcal{N}$(1.9,0.2), from which we derived $a_{\rm b}/R_{\star}$ at each step of the MC sampling. We recover the prior distribution in the posterior of $\rho_{*}$.}
      }
      \end{center}
\end{table*}

\section{A new low mass companion at close separation detected with SPHERE} \label{sec:bd}

\begin{figure}
    \centering
    \includegraphics[width=0.5\textwidth]{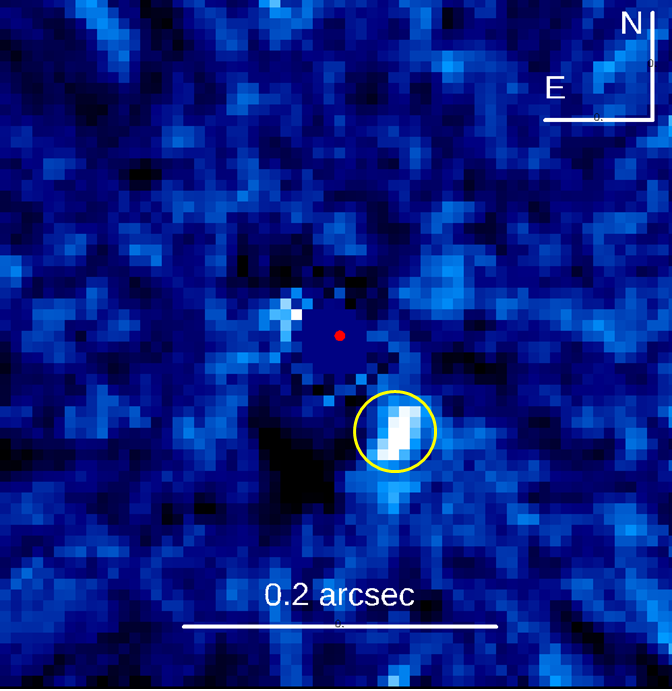}
    \caption{Detection of the low-mass companion in the non-coronagraphic SPHERE images (average of the two epochs). The red dot marks the position of the star and the yellow circle shows the companion. The star image has been subtracted using the differential approach described in the text. The small elongation of the detected source is due to the subtraction between the images taken at different angles, in a similar way to the classic ADI pattern. }
    \label{fig:sph_comp}
\end{figure}

A close companion is clearly identified at S/N=14 in the non-coronagraphic SPHERE data sets obtained at both epochs (2019 Oct 11 and 2019 Dec 18) at a projected separation of 84.5$\pm$3.6 mas and position angle 212.0$\pm$1.5 deg  (Fig. \ref{fig:sph_comp}); this is the average value obtained at the two epochs, that are fairly close in time, so that we do not expect significant orbital motion between them. 
The magnitude differences with respect to HD 18599 result  $6.42\pm 0.07$~mag in $J$ band and $6.05\pm 0.07$~mag in $H$ band\footnote{This $H$ band actually only includes the wavelength range accessible to the SPHERE IFS between 1.50 and 1.65~$\mu$m and it is then not coincident with the usual definition.} (internal errors only). 
These values correspond to absolute magnitudes of $M_J=10.91\pm 0.07$\ and $M_H=10.14\pm 0.07$~mag. For the adopted age of $400\pm 100$~Myr, the application of \citet{baraffe2015} models yields a mass of $83^{+4}_{-6}$~M$_J$. 
The observed J-H color (0.77$\pm$0.10) implies a spectral type between very late M to early L, in agreement (although less constraining in terms of mass) to the estimate provided above from the absolute magnitudes. 

The object is not detected in the NaCo images, as expected from
the contrast limits shown in Fig.~\ref{fig:magcontrastnaco}. For the same reason, it remains undetected in shallower imaging of TOI-179 published by \citet{ziegler2020}.

The physical association with the central star is not fully demonstrated, as the time baseline between the two SPHERE observations is too short for the common proper motion test.
However, we note that a stationary background object would be expected to lie at a projected separation of 203 mas from the central star at the epoch of the NaCo observations, above our detection limits (Fig.~\ref{fig:magcontrastnaco}) for reasonable H-L colors of the object\footnote{Note that a field object with blue colors is ruled out by the observed colors measured on SPHERE data.}. This does not rule out a non-stationary background object with intrinsic proper motion \citep{nielsen2017} keeping the projected separation smaller in 2016, but this option has a very low probability to occur.

Considering also the existence of the RV trend and astrometric signature (see below), which are compatible with the source detected in imaging,
and the observed very red color, we consider in the following the object as a physical companion, that
results to be a very low mass star or a high-mass brown dwarf, mostly depending on the stellar age. We label it HD 18599 B\footnote{The wide companions CD-56 593A and B are not identified as HD 18599 B and C in SIMBAD at the time of writing, although \citet{mugrauer2020} label the unresolved entry (from Gaia DR2) as TOI-179BC. WDS instead labels CD-56 593A and B as WDS 02570-5613A and B, respectively, and HD 18599 as WDS 02570-5613C, in spite of the latter being the brightest and most massive component in the system. We propose to identify the CD-56 593 components as HD 18599 C and D to avoid ambiguities.}.

\subsection{Origin of the long-term RV trend of TOI-179}
\label{sec:trend}

\begin{figure}
    \centering
    \includegraphics[width=0.5\textwidth]{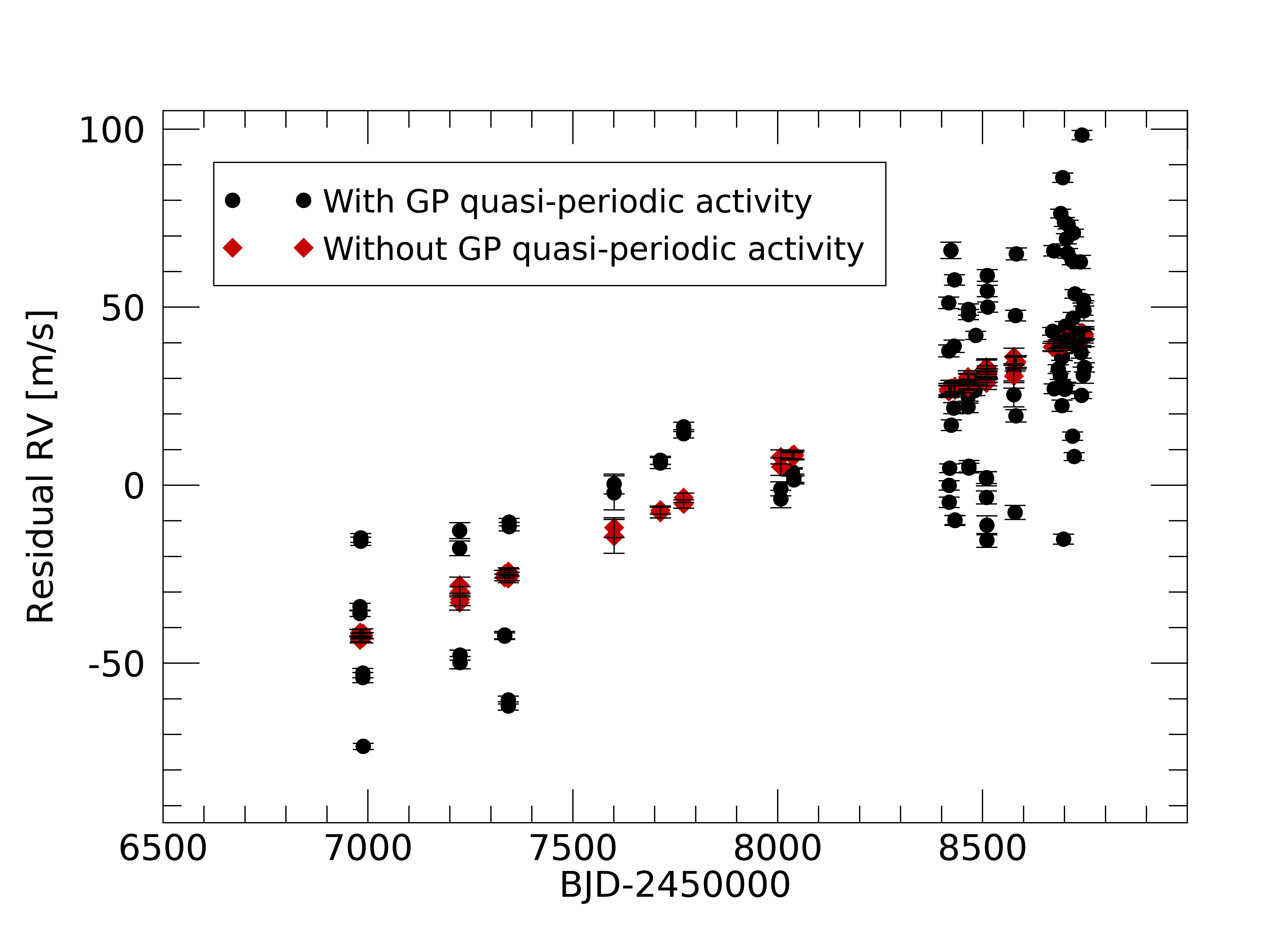}
    \caption{Residuals of the HARPS RVs. Black dots: after removing the best-fit spectroscopic orbit of TOI-179~b while keeping the activity signal. Red diamonds: after removing also the GP quasi-periodic activity signal. }
    \label{fig:rv_resid}
\end{figure}

As shown in Sect. \ref{sec:rv_tess_fit_toi179}, the RV time series shows a clear long-term trend with a slope of 17.5$\pm$2.9 $\ms $yr$^{-1}$. Fig. \ref{fig:rv_resid} shows the RV residuals, after removing the signal due to TOI-179~b, with and without the quasi-periodic activity signal modulated over the stellar rotation. The inclusion of a quadratic term is not highly significant (Table \ref{t:toi-179_pl_param}).

We noted that the activity indices, in particular the H${\alpha}$ index measurements also show a possible long-term trend (Sect. \ref{sec:activity}). While it is possible that this is also contributing to some extent to the long-term signal seen in the RV time series, we noticed that the presence of the companion revealed in imaging and of the astrometric signature (Sect. \ref{sec:astrometry}), and the amplitude itself of the RV trend, much larger than the typical RV signatures of activity cycles \citep{lovis2011,carolo14} strongly argue for a Keplerian origin of the observed RV trend. Moreover, this assumption is supported by the lack of a long-term trend in the BIS diagnostic (Fig. \ref{fig:activity}, lower-right panel), which instead shows the anti-correlation with the RVs typically observed in active, late-type stars and significant periodicity corresponding to the rotational modulations.

Nonetheless, we quantified a possible overestimate on the measured RV acceleration $\dot{\gamma}$ due to the long-term activity by repeating the fit described in Sect. \ref{sec:rv_tess_fit_toi179}, carrying out two tests using the spectroscopic activity diagnostics based on the H-alpha and Calcium lines (Sect. \ref{sec:activity}). In the first test, we included a linear term with the \textit{Ha16} index in the model to correct for a RV-activity diagnostic correlation. We used large and uninformative priors for the coefficients of the linear model, and we got $\dot{\gamma} = 0.042^{+0.009}_{-0.008} \ms d^{-1}$, slightly less but in agreement with the previous estimate  ($0.048\pm0.008 \ms d^{-1}$), without changes in the values planetary parameters. 

In a second test, we considered a more complex model, consisitng of 28 free parameters, including the time series of the S index. The model includes a linear term with the S-index to correct for a RV-activity diagnostic correlation, and, since the rotational modulation is detected in the S index periodogram (Sect. \ref{sec:activity}), we also include a GP quasi-periodic correlated term to model the S index time series, together with a quadratic long-term trend. Except for the scale amplitude \textit{h} of the GP, the three hyper-parameters left ($\lambda$, $w$, and $\theta$) are shared between the activity diagnostic and RVs. We found $\dot{\gamma} = 0.047^{+0.008}_{-0.009} \ms d^{-1}$, in agreement with the previous estimate, without changes in the values of the planetary model parameters. We note that the best-fit values of the stellar rotation period $\theta$ and that of the evolutionary activity time-scale $\lambda$ are perfectly consistent with those obtained by fitting the RVs alone, showing that a GP quasi-periodic trained on the S index does not improve or even change the best-fit solution for the activity signal present in the RVs. 
Hence we conclude that the observed long-term RV trend is mostly of Keplerian origin and assume in the following that it is due to HD 18599 B. Finally, the trend can not be due to secular acceleration \citep{zechmeister2009}, as this effect is removed by the TERRA pipeline when deriving the RVs.

\subsection{Astrometric signatures} \label{sec:astrometry}

\citet{kervella2022} detected a rather large proper motion anomaly (PMA) of $0.406\pm 0.036$~mas/yr at a position angle of 265.54 degree by comparing the proper motion measured by Gaia eDR3 with the long-term proper motion obtained comparing the position of TOI-179 at the Gaia eDR3 epoch (J2016.0) with that obtained at the Hipparcos epoch (J1991.25). This large PMA (detected with a highly significant S/N of 11.4) is clearly indicative of an acceleration due to a companion. This cannot be attributed either to the transiting planet or to the wide companions CD-56 593 A and B; it may rather be ascribed to the low mass companion observed by SPHERE (HD~18599~B), further supporting the high detection efficiency for this kind of objects around stars with significant PMA \citep{bonavita2022}. We note that the PMA measured at the Hipparcos epoch are $-0.511\pm 0.720$ and $-0.301\pm 0.690$~mas/yr in right ascension and declination, respectively. These values are not significant, but comparable with the result obtained at the Gaia epoch.

\subsection{Constraints on the orbit of HD 18599~B} \label{sec:orbit}

We used the astrometric epochs obtained with SPHERE, the RV long-term trend, and the proper motion anomaly signatures at both Hipparcos and Gaia epoch to constrain the orbit of HD~18599~B. 

As a first qualitative guess, we infer that the object is caught at a projected separation much smaller than the physical semimajor axis.
Indeed, a strong RV signal (K $\sim$  1 km/s for edge-on orbit) of period $\sim$3.6 yr is expected for a companion with the semimajor axis equal to the 3.3 au projected separation of HD 18599 B. Instead we observe a monotonic behaviour of the RVs over 4.8 yr (upward trend), with no significant power at 3-5 years in the periodogram. A very small RV amplitude can be explained with a pole-on orbit, but in this case the astrometric motion should be roughly perpendicular to the projected separation, which is very different from the observed Gaia proper motion anomaly.

For a quantitative analysis, we first used the program ORBIT \citep{Tokovinin2016}  \footnote{\url{https://zenodo.org/record/61119\#.Xg83GxvSJ24}   } 
as a tool to find possible orbital solutions compatible with the observational data, fixing the total mass for the system obtained considering the mass of the star and of the companion derived from photometry. We explored the period range from 3
to 150 yr (the small apparent separation of the SPHERE detections makes longer periods unprobable). While data are not enough to completely define this orbit, various constraints can be obtained. In general, we found two families of orbital solutions compatible with the observational data, either with periods shorter than about 30-40 yr, or with periods longer than 60-80 yr, while orbits with intermediate periods (roughly twice the separation between Hipparcos and Gaia observations) are excluded. 

In the case of short periods, the small separation of the companion detected by SPHERE and the shallow almost linear trend on the RV curve over a quite long range of five years point towards a highly eccentric orbit ($e\gtrsim 0.9$) seen at high inclination ($i\sim 100$ degree) and with an angle between the periastron and the line of nodes $\omega$ not too far from 90 degrees (the longer the period, the closer is this condition). The mass for HD~18599~B should be small (in the range between 20 and 40~$M_{\rm Jupiter}$) that is essentially a consequence of the observed shallow trend in radial velocities. This is incompatible with the value derived from photometry. 

On the other hand, orbits belonging to the long period family have moderate eccentricity ($e\sim 0.5$), an even higher inclination ($i\sim 95$ degree), and a wider range of values for $\omega$. The mass of HD~18599~B found with these orbits are in the range between 60 and 120~$M_{\rm Jupiter}$, compatible with the value indicated by photometry. If we rather use this last as a prior, the orbital period is constrained to be not too far from 100 yr.

The two orbit families also correspond to very different values for the periastron distance; while the short period orbits have very short periastron - around or less than 1 au - the long period orbits have much larger values in the range between 7 and 14 au. This leads to a very different strength of the dynamical interaction between the secondary and the transiting planet. Indeed, we notice that for the orbit of the transiting companion to be stable, the periastron distance for HD~18599~B should be larger than $\sim 0.24$~au, requiring that the eccentricity should be smaller than 0.97 for short period orbit, and a bit larger for long period ones. 

Finally, the very high value of the inclination suggests that the orbit of HD~18599~B might be nearly co-planar with that of the transiting planet.

\begin{figure*}
\centering
\includegraphics[width=17truecm,angle=0]{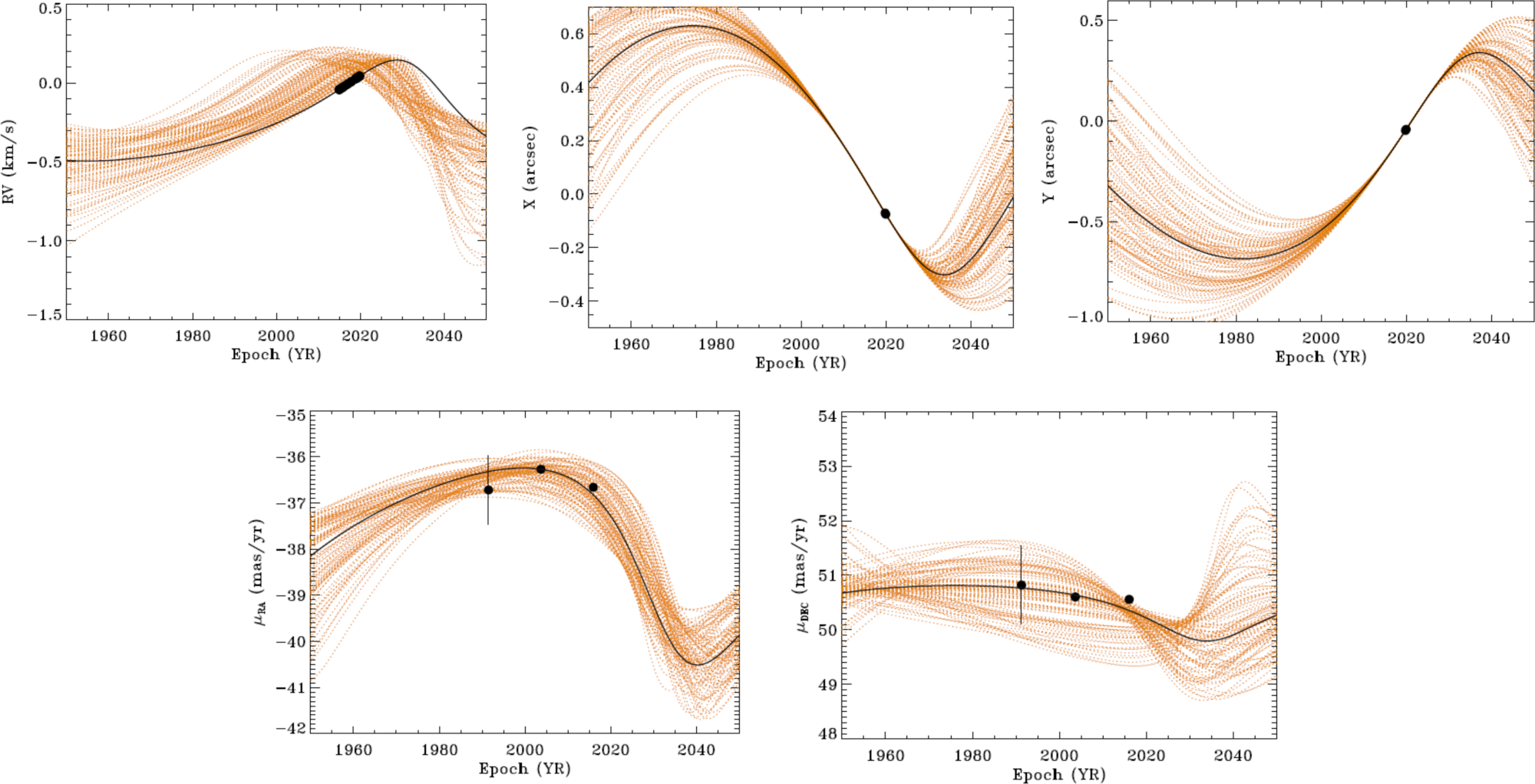}
\caption{\label{fig:orbit118} 
RV, sky position, and  motion on the sky for the long-period solution as resulting from the MCMC described in the text. The black line
shows the nominal best fit solution. The dashed orange lines represent a random selection of orbital solutions drawn from the posterior distributions of the model parameters from the DE-MCMC analysis. Black dots are the observational data from HARPS (RV panel), SPHERE (X and Y panels) Gaia and Hipparcos (proper motion panels).} 
\end{figure*}

Next, we explored the parameter space of possible solutions using a Bayesian analysis of the combined RV, absolute and relative astrometry observations using a differential evolution Markov chain Monte Carlo (DE-MCMC) method \citep{TerBraak2006,Eastman2013}. As described in detail in \citet{Drimmel2021}, the combination of the three types of datasets allows to directly determine the mass ratio $q$. By placing narrower and broader uniform priors on the orbital period 
($P$ in range [3., 60.] yr and [3., 200.] yr, respectively)
we reproduced in general terms the solutions obtained with the ORBIT package. By adopting an additional, indirect prior on the system distance based on the knowledge of the Gaia eDR3 parallax \citep{BailerJones2021}, the shorter-period solution converged (based on the Gelman-Rubin statistics, see e.g. \citealt{Ford2006}) to 
$P \sim 45 $ yr, $e \sim 0.89 $, $q \sim 0.092$.
The derived inclination is found to be $i \sim 85$ deg.    
Given the primary mass reported in Table \ref{t:star_param}, the inferred companion mass is $M_B \sim 82$ M$_J$. 
The equivalent set of values for the parameters of the longer-period solution is: $P \sim 119$ yr, $e \sim 0.32$, $q \sim 0.059$, $i \sim 73$ deg, and $M_B \sim 53$ M$_J$. 
The long-period solution is represented grafically in Fig. \ref{fig:orbit118}.

It should be noted that the long-period solution is clearly favored in terms of likelihood ($\Delta\ln \mathcal{L}>10$). Nevertheless, it is also clear that the effectively single epoch for imaging data and the long-term trend in RVs make any robust inference on the exact orbital configuration  and mass of HD 18599 B unfeasible: the above numbers should be considered simply representative of the two broad families of orbits, and corresponding approximate companion mass, that are allowed given the observational constraints at hand: {\it a)} a shorter period, very high-eccentricity, almost edge-on configuration,
and {\it b)} a longer period, lower-eccentricity, still highly inclined configuration. 

\subsection{Ruling out HD~18599~B as host of transit}
\label{sec:notransit}

The presence of a close companion around TOI-179 calls for a check of the possibility that the observed transits are not occurring on the central star and that HD 18599~B is a diluted eclipsing binary \citep[see, e.g.][]{torres2004}.
This hypothesis is dismissed for a number of reasons.
On one hand, the expected magnitude difference between TOI-179 and HD 18599 B in V band
is more than 12 mag \citep[using the models by ][]{baraffe2015}, making the contamination of the HARPS spectra negligible. 
Therefore, the presence of a Doppler signal with the same period and phase of the photometric can not be ascribed to HD 18599 B.
On the other hand, the photometric properties of the transits (depth, duration, and shape)
are also not compatible with diluted eclipses occurring on the low mass companion.

\section{Discussion}
\label{sec:discussion}

\subsection{Architecture of TOI-179 system}

The architecture of the TOI-179, with a close-in, low-mass transiting planet in eccentric orbit and an outer companion with mass close to stellar/substellar boundary, possibly coplanar with the transiting planet, is fairly unusual in the wide zoo of known planetary systems.

Few examples are found of low-mass planets in close orbits with outermost companions (brown dwarfs or very low mass stars) within few tens of au. The M dwarf GJ 229 is well known for hosting the first T type companion discovered by \citet{nakajima1995}. More recently, two low-mass planets candidates were claimed through RVs by \citet{tuomi2014}, one of which (GJ 229 c, $m \sin i$ 7 $M_{\oplus}$, period 122 d, eccentricity 0.19) was confirmed with further observations by   \citet{feng2020}. As this planet is non-transiting, the relative inclination with respect to the BD orbit is unknown. The orbit of the BD has a semimajor axis of 34.7 au and is highly eccentric \citep[0.846$\pm$0.015, ][]{brandt2020}. 
Among the planet hosts with very low-mass stellar companions (mass $\le 0.1 M_{\odot}$) at close
separation ($\le 100 $ au) we mention only three cases listed in catalogs by
\citet{thebault2015}
\footnote{updated version at \url{http://exoplanet.eu/planets_binary/}}
and \citet{schwarz2016} \footnote{updated version at \url{https://www.univie.ac.at/adg/schwarz/multiple.html}} .
The G dwarf GJ 3021 has a 0.09 $M_{\odot}$ star
at 68 au (projected separation) \citep{chauvin2006} and a moderately massive giant planet ($m \sin i$ 3.4 M$_{J}$) at 0.5 au. 
K2-126 is a K7V Hyades member with three low-mass transiting planets \citep{mann2018} and a 0.10 $M_{\odot}$ stellar companion at a projected separation of 40 au \citep{ciardi2018}.
Finally, HD 42936 is an exceptionally compact system, formed by a low-mass planet, possibly evaporating, very close to the K0 central star and a 0.076 $M_{\odot}$ star (minimum mass from RV solution) with semimajor axis 1.22 au and eccentricity 0.59 \citep{barnes2020}. Thus, the stability zone around the primary extends only up to 0.16 au.
This architecture is qualitatively similar but more extreme than the high eccentricity case for the TOI-179 system described above.

TOI-179 then represents one of the few known cases of a star hosting both at least a planet and a second companion (a very low mass star or BD) in close orbit, making it an interesting laboratory for our understanding of formation of both these kinds of objects in the same system. The characteristics of TOI-179b (see also below) suggests a formation path through the core-accretion mechanism \citep[e.g.,][]{mordasini2012}, while HD18599 B might be the outcome of disk  fragmentation \citep{stamatellos2009}.
The likely coplanar orbits and eccentric  orbit for the outer companion
add further constraints on the processes of formation and evolution, although the ambiguities on the orbit of HD 18599B do not allow conclusive inferences. 

The presence of the other two companions CD-56 593 A and B should be mentioned as a potential source of dynamical interactions in the system, although their current projected separation (3400 au) is very  large. Rearrangement of the configuration of the system may have occurred in the past \citep[see, e.g.,][]{marzari2007,kaib2013}, or the orbit may be highly eccentric, although these hypothesis remain still speculative.

\subsection{Constraints on additional companions} \label{sec:addcomp}

Photometric, spectroscopic, and AO data do not provide indications for additional
companions beside the transiting planet TOI-179b and the imaged low-mass object HD 18599 B. 

\begin{figure}[!hb]
\centering
\includegraphics[width=\columnwidth,angle=0]{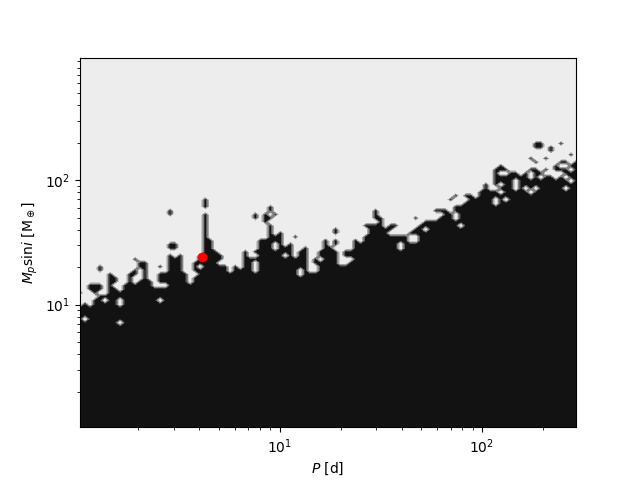}
\caption{\label{fig:detlim} 
Detection function map of the RV time series of TOI-179. The white part corresponds to the area in the period - minimum mass space where additional signals could be detected if present in the data, while the black region corresponds to the area where the detection probability is negligible. The red circle marks the position in the parameter space of the transiting planet TOI-179b.}
\end{figure}

More in detail, from the RV time series, we can derive the detection limits for additional RV  objects, following the Bayesian technique described in \citet{pinamonti2022} , adopting the results from the joint modelling of RVs and photometric transits as priors on the TOI-179b planetary signal. The resulting detection function map is shown in Fig. \ref{fig:detlim}.
We can rule out the existence of planetary companions with $m \sin i$ larger than 28 and 80 M$_{\oplus}$ with orbital periods of about 10 and 100 days, respectively.

\begin{figure}[htbp]
\centering
\includegraphics[width=\columnwidth]{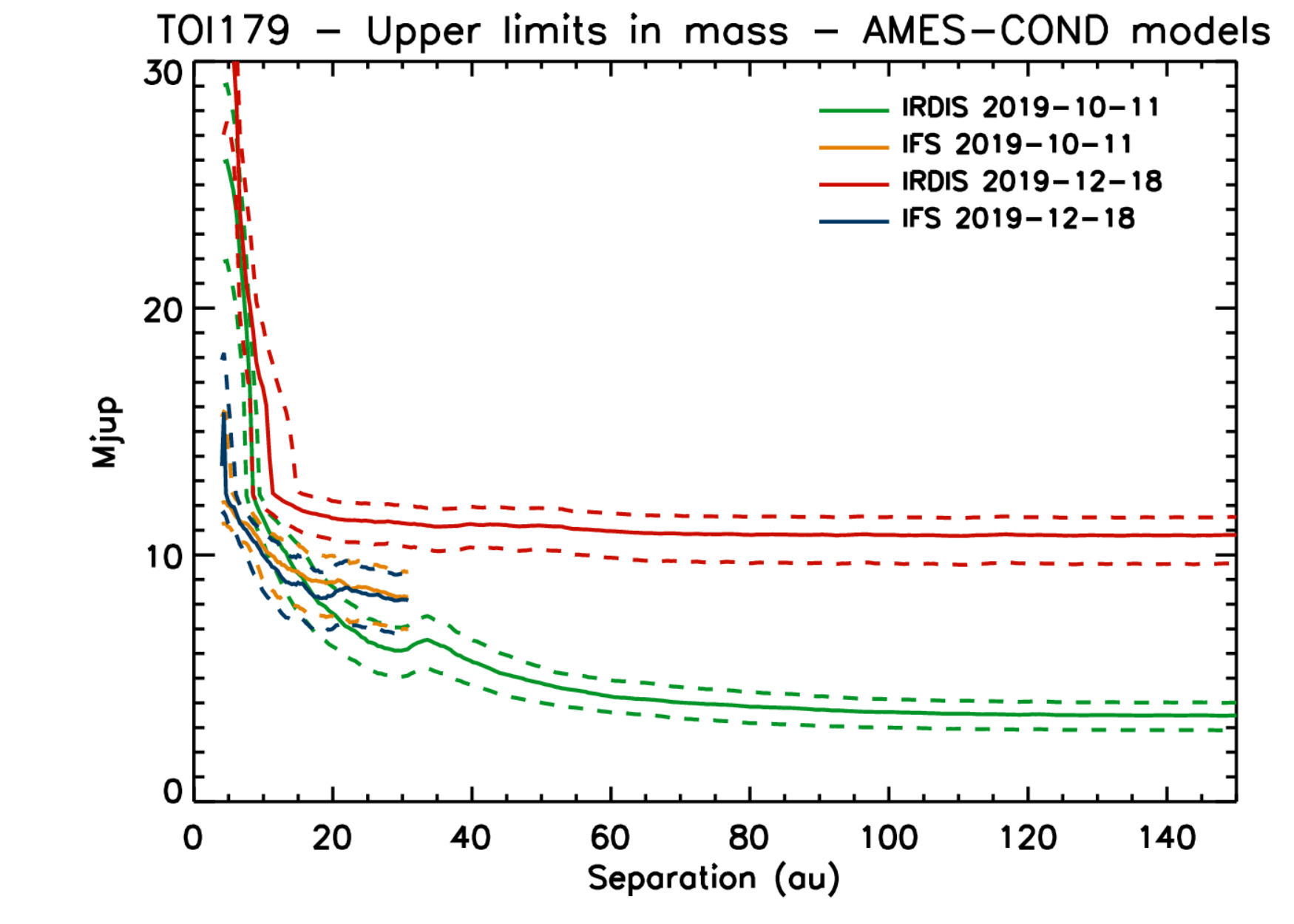}
\caption{\label{fig:masslimit} Mass detection limits expressed in $M_{Jup}$ as a function of the separation for both epochs and both for IRDIS and IFS. The dashed lines represent the possible mass ranges due to the age uncertainty.}
\end{figure}

We also used the coronagraphic data described in Section~\ref{s:spheredata} to obtain a final image of the FoV around 
TOI-179. No obvious point source was identified both in the IFS and in the IRDIS FoV. 
We used the contrast limits derived in Sect. \ref{s:spheredata} and shown in Fig. \ref{fig:magcontrast}  
to define the mass limits for both the epochs and for both the SPHERE instruments assuming an age 
of 400$\pm$100~Myr and using the AMES-COND models 
\citep{allard03}. The results of this procedure are displayed in Fig.~\ref{fig:masslimit} where we can see that at 
separations from the host star lower than $\sim$20~au we can
exclude companions with a mass of the order of 10~$M_{Jup}$ while
at separations larger than 50~au we can exclude companions 
with masses larger than 3-4~$M_{Jup}$. 
The lack of additional companions is not surprising considering the presence of HD 18599B at a projected separation much smaller than the true semimajor axis, with indications for an eccentric orbit.

\begin{figure}[htbp]
\centering
\includegraphics[width=\columnwidth]{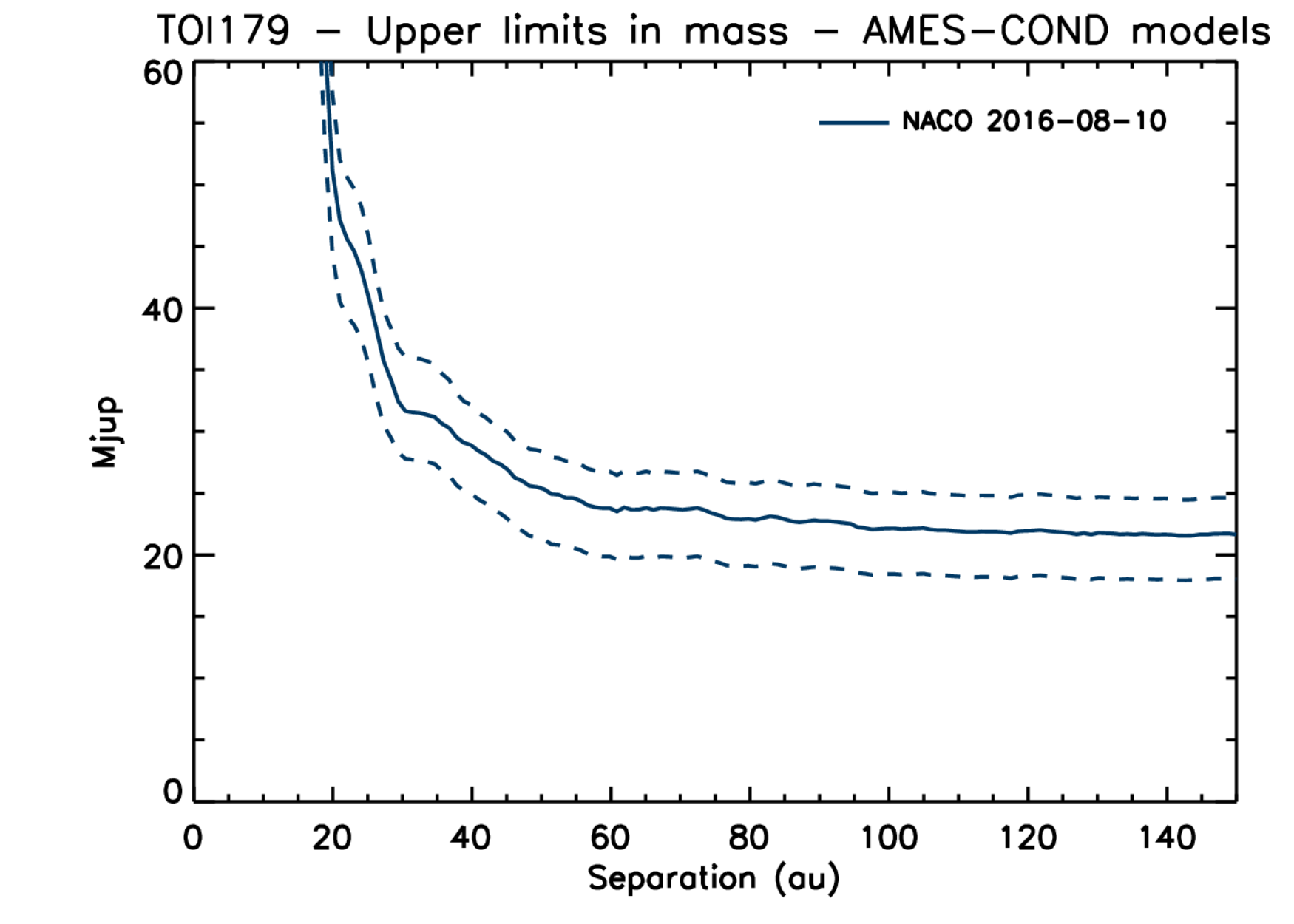}
\caption{\label{fig:masslimitnaco} Mass detection limits expressed in $M_{Jup}$ as a function of the separation obtained from the NACO data. The dashed lines represents the possible mass ranges due to the age uncertainty.}
\end{figure}

The same procedure was also performed to infer the mass detection limits from the  NaCo data (Sect.~\ref{sec:naco}). The results are displayed in  Fig.~\ref{fig:masslimitnaco}. 
As expected the mass limits that we obtain in this case are much higher than those obtained with SPHERE.

\subsection{Dynamical interactions}
\label{sec:dyn}

While the lack of a well-defined orbit for the outer companion does not allow us to address in depth the occurrence of dynamical interactions between the objects in the system, we performed some explorative evaluations, considering representative solutions for the two families of possible orbits identified in Sect. \ref{sec:orbit}.

The close distance of the planet to the primary star ensures long-term stability to the system in both configurations in spite of the presence of the low-mass companion. Due to the plausible high value of eccentricity of the massive companion and of the planet it is not possible to apply the Laplace-Lagrange secular theory. The amplitude of the eccentricity oscillations of the planet has been numerically evaluated to be of the order of 0.1-0.2 for the first configuration with a period of $\sim 10^5$ yrs, depending on the initial values of the orbital angles, and of the order of 0.04-0.08 for the second configuration with periods of the order of $\sim 10^7$ yrs.  As a consequence,  the high eccentricity value of the planet is not due to the secular perturbations but it is potentially related to an earlier phase of dynamical instability (planet-planet scattering). The Kozai mechanism appears unlikely considering  the probable configuration close to coplanarity of the planet and low-mass companion. It is possible that a second planet was present in the system and it was scattered out of the system leaving the observed planet in an excited orbit.  

In the first configuration, the fast secular oscillations may prevent the tidal circularization of the orbit while the second configuration appears to be more responsive to tidal forces due to the smaller secular oscillations and their longer period. 

The high eccentricity of HD 18599 B in the first configuration may not be primordial and even in this case a chaotic origin may be invoked. A period of dynamical instability involving the outer stars may have driven the object closer to the primary after the scattering of an additional star.

\subsection{Tidal evolution and the origin of the eccentricity of TOI-179b}
\label{sec:tides}

The moderately large eccentricity of TOI-179b (0.34$^{+0.07}_{-0.09}$) is unusual among planets with measured eccentricities. \citet{correia2020} found that typically planets with radii between 3 to 9 R$_\oplus$ have small but significant eccentricities (typically $\sim$ 0.15). High eccentricities appear instead less common for planets below 3 R$_\oplus$, although data are rather sparse. The only other known transiting planet with radius $<$ 9 R$_\oplus$, period shorter than 5 days, and eccentricity larger than 0.3 is TOI-942b \citep{carleo2021}, although the measurement suffers of large uncertainty. The case of TOI-942b however appears different, as there is another close-in, low-mass, transiting planet TOI-942c with a period of 10.2 d, while TOI-179 has a more massive companion at the BD/low-mass star boundary at much wider separation. Furthermore, the star TOI-942 is younger (50 Myr).

In order to shed light on the origin of the moderately high eccentricity, we estimated the timescale of tidal circularization, that depends mainly on the tidal modified quality factor $Q^{\prime}_{\rm p}$ of the planet, the role of tidal dissipation inside the star being much less important, even in the case we assume a dynamical tide produced by inertial waves during most of the past evolution of the system, given that the rotation period of the host star is close to twice the orbital period \citep{Ogilvie14}. The value of $Q^{\prime}_{\rm p}$ is highly uncertain because it depends on the internal stratification and composition of the planet. Assuming that tidal dissipation occurs mainly in a rocky/icy structure, $Q^{\prime}_{\rm p} \approx 10^{3}$ \citep[see][]{Tobieetal19,Lanza21} and the circularization timescale is of the order of $\approx 50$~Myr.  Nevertheless, given the uncertainty about the internal structure of the planet and consequently on its rheology, it is important not to overinterpret this estimate. An upper limit to the circularization timescale may be found by assuming $Q_{\rm p}^{\prime} \approx 10^{5}$ that is characteristic of a giant planet where most of the tidal dissipation occurs in fluid layers \citep[][Sect.~5.4]{Ogilvie14}, that is likely not the case for TOI-179b. In this case, the eccentricity of TOI-179b can be a remnant of its formation and migration processes because the estimated circularization timescale is $\sim 4$~Gyr. All these estimates are based on the constant time lag tidal model of \citet{Leconteetal10}, where the transformation between the modified tidal quality factor and the time lag is made through their Eq.~(19).

On the other hand, the low mass companion HD 18599 B might also be involved, pumping the eccentricity of the transiting planet through dynamical interactions, as discussed in Sect. \ref{sec:dyn}.

Independently of its origin,
the eccentric orbit is expected to lead to a pseudosynchronous rotation of the planet with a rotation period $\sim 1.7$ times shorter than the orbital period and an internal tidal dissipation $P_{\rm diss}$ that ranges from $2\times 10^{18}$ to $2 \times 10^{20}$~W for $e=0.34$ and $Q^{\prime}_{\rm p}$ ranging from $10^{5}$ to $10^{3}$  \citep[$P_{\rm diss}$ is directly proportional to $e^{2}$ and inversely proportional to $Q^{\prime}_{\rm p}$; see][]{Milleretal09}.   The estimated surface heat flux ranges between $\approx 150$ and $\approx 5 \times 10^{4}$~W~m$^{-2}$ that is hundreds or thousands of times larger than in the case of the Jovian moon Io, suggesting that large scale and intense volcanic activity can take place on TOI-179b. Its observability largely depends on the structure of its outer layers and the presence of an atmosphere, but certainly this is an interesting candidate where to look for signs of volcanic plumes through transit observations.

\subsection{Planet density: comparison with other transiting planets}
\label{sec:density}

\begin{figure}[htbp]
\centering
\includegraphics[trim={2.5cm 2.9cm 2.5cm 2.9cm}, width=0.65\columnwidth, angle=270]{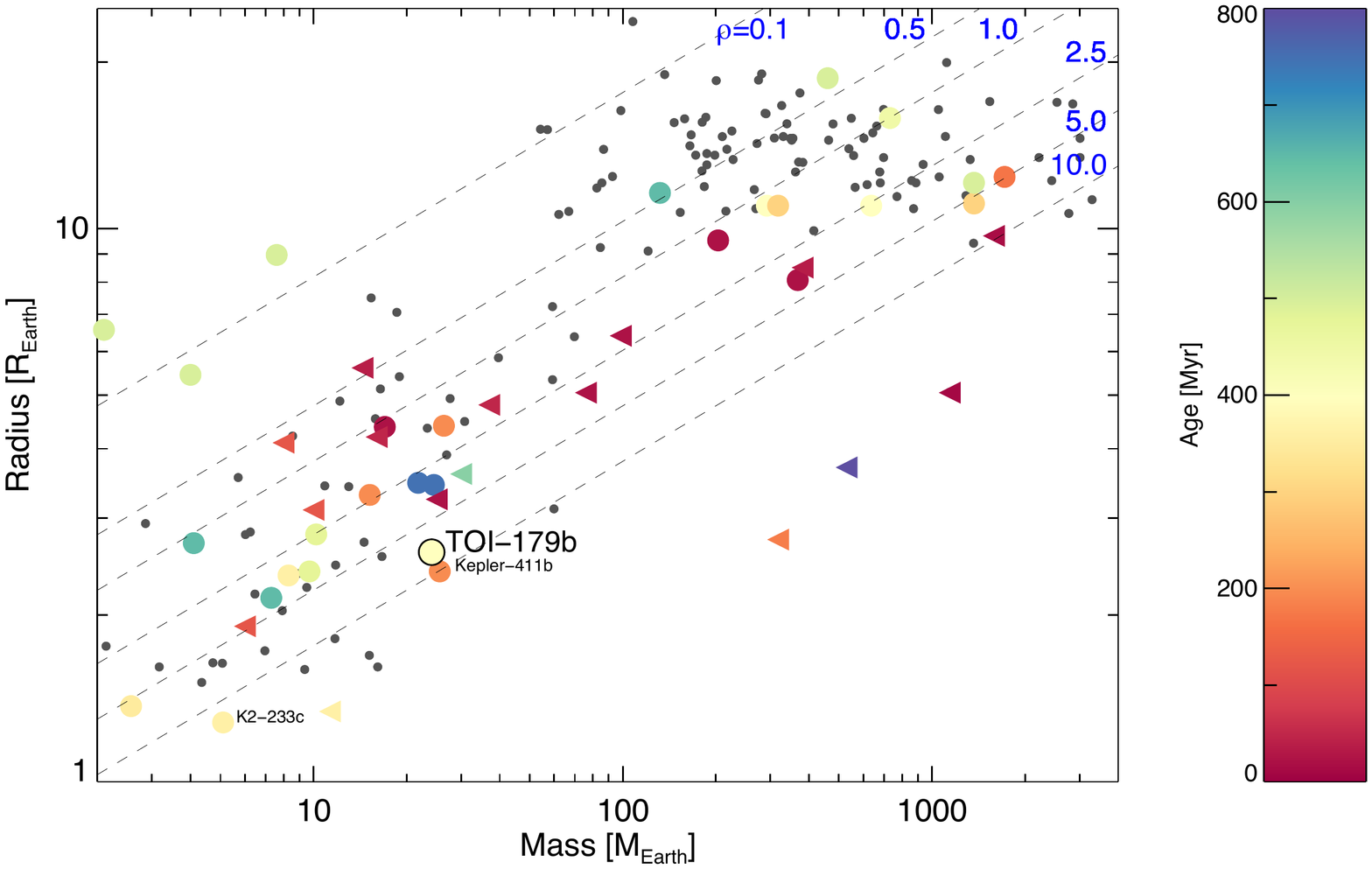}
\caption{\label{fig:mr}  \textit{Upper panel}: Mass-Radius diagram of the well characterized exoplanet population (grey dots) and of the young transiting planets as a function of the stellar age (colour code on the right). Coloured circles indicate planets with measured mass while triangles indicate planets with mass upper limit only. Dashed lines represent the loci of equal density [g cm$^{-3}$].  }
\end{figure}

Aiming to put the high-density planet TOI-179 b in the context, we selected all  planets with masses and radii measured with uncertainty better than 20\% and 10\%, respectively. We used the Exo-MerCat tool\footnote{User interface available at \url{https://gitlab.com/eleonoraalei/exo-mercat-gui}} \citep{alei2020} to retrieve the planetary parameters from the four main exoplanets archives\footnote{Exoplanets Encyclopaedia (\url{http://exoplanets.eu/}), the NASA Exoplanet Archive (\url{https://exoplanetarchive.ipac.caltech.edu/}), the Open Exoplanet Catalogue (\url{http://www.openexoplanetcatalog.com/}) and The Exoplanet Data Explorer (\url{http://exoplanets.org/})} and produce the Mass-Radius (M-R) diagram in  Fig. \ref{fig:mr}. The grey dots represent the known planet population, mainly composed by mature systems. We also add the young transiting exoplanets known to date (age < 800 Myr) represented by coloured dots according to the age in the colorbar on the right side of the plot. Because of the high level of the stellar activity (hampering the recovery of the planetary signal) or because of the intrinsic difficulty to perform a proper RV follow-up (e.g. very faint stellar hosts), a robust determination of the planetary mass is not available for a significant fraction of those planets. These cases are depicted with coloured triangles pointing toward lower values, since we only know the mass upper limits. Finally, the dashed lines indicate the location of equal planet density, as marked at the top of the figure. 

Fig. \ref{fig:mr} shows that TOI-179 b lies on the denser edge of the super-Earths/Neptune-like planets M-R distribution, at least among the planetary companions younger than $\sim 800$ Myr with measured mass. It has similar physical properties to those of Kepler-411~b, including a comparable orbital period of 3 days. The latter planet orbits a young ($\sim 200$ Myr) K2V star with an external M dwarf companion, which also denotes similarity between the stellar environments. On the other hand, Kepler-411~b belongs to a fairly compact four-planet system \citep{sun2019}, indicating  different formation and evolution paths.
Among the well characterised young/intermediate age planets also K2-233~c has a high density (see Fig. \ref{fig:mr}), placing it in the family of pure iron exoplanets \citep{lillobox2020}. Also K2-233~c orbits a K3V star with age comparable to TOI-179, but like the case of Kepler-411, it belongs to a multi-planet system.

\begin{figure}[htbp]
\centering
\includegraphics[trim={2.5cm 2.9cm 2.5cm 2.9cm}, width=0.65\columnwidth, angle=270]{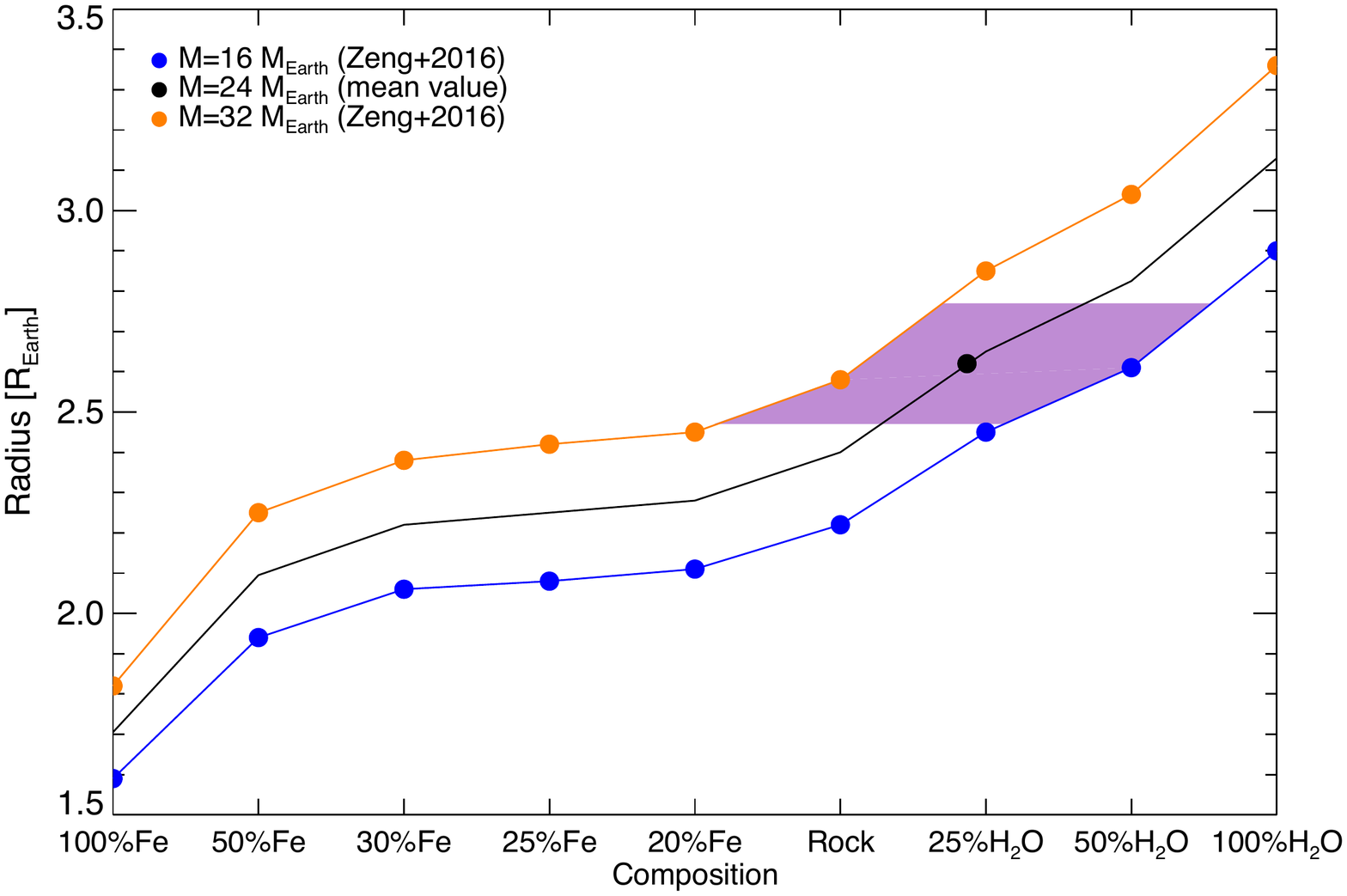}
\caption{\label{fig:models} Models of planetary composition reported from \cite{zeng+2016} for planets with 16 and 32 M$_{\oplus}$ (blue and orange trace, respectively). The black dot represents the putative location of TOI-179 b by considering a mean model between the considered models. The purple area indicates the range of compositions describing the structure of TOI-179 b. }
\end{figure}

To investigate the possible composition of TOI-179~b, we considered Table 2 in \cite{zeng+2016} where models for planets with mass equal to 16 (blue trace in Fig. \ref{fig:models}) and 32 M$_{\oplus}$ (orange trace) are reported. The traces describe the radius variation according to the possible planet composition. Since the measured mass for TOI-179 b is $\sim 24$ M$_{\oplus}$, we averaged the values reported by \cite{zeng+2016} to produce the model depicted with a black line and locate our target at the nominal value of the radius (black dot). Considering the error bars for both the mass and the radius, the planetary composition is described by the purple area in Fig.~\ref{fig:models}, ranging from pure rock with a small fraction of iron to 25\% rock + 75\% water, with typical structure showing 75\% rock + 25\% water.  

\begin{figure}[htbp]
\centering
\includegraphics[trim={1.5cm 2.9cm 2.5cm 2.9cm}, width=0.7\columnwidth, angle=270]{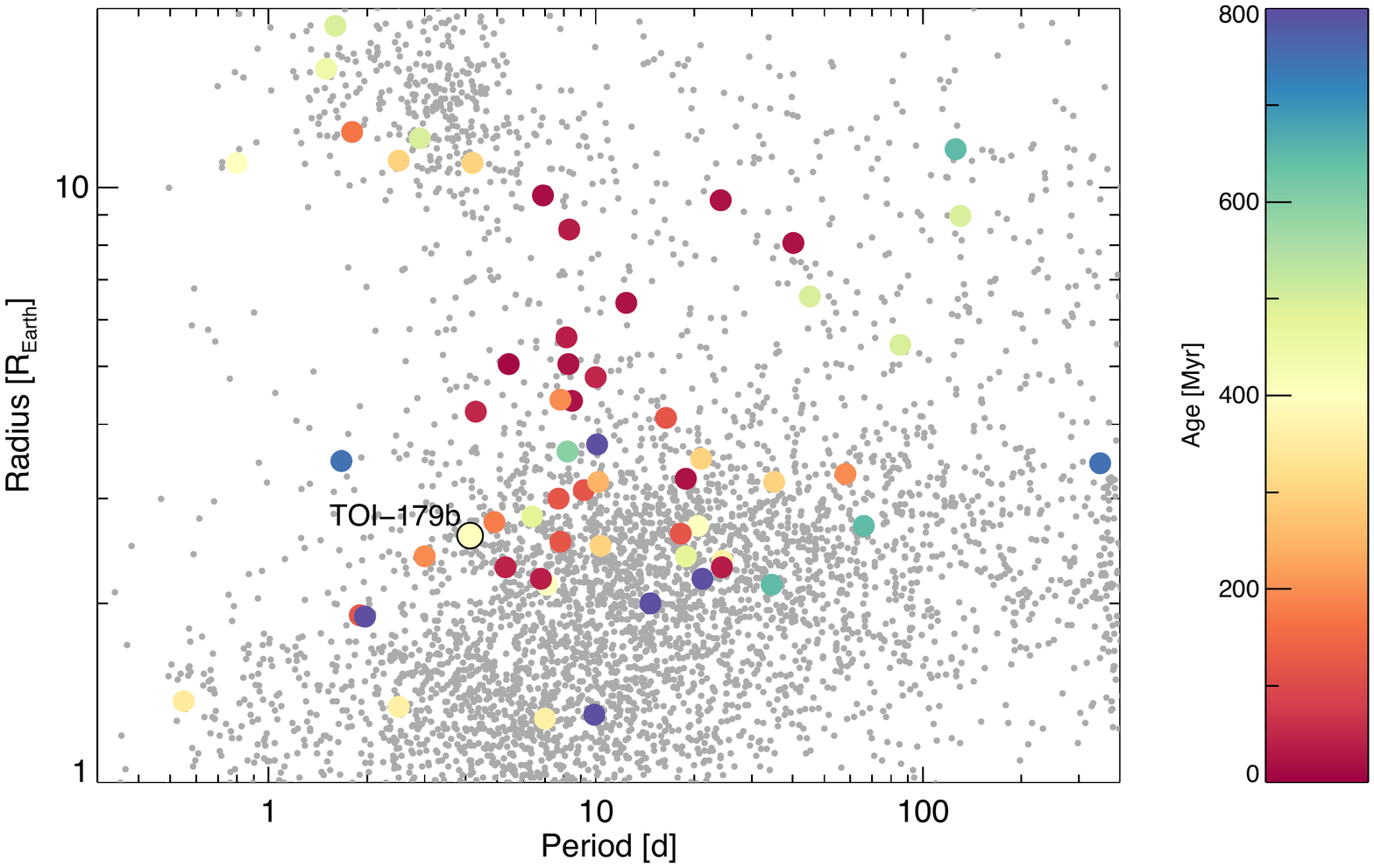}
\caption{\label{fig:pr} Period-radius distribution of the transiting planets  with radius known better than 10\% (grey dots). Planets younger than 800 Myr are depicted with different colors according to their age as defined in the colorbar. }
\end{figure}

Finally, Fig.~\ref{fig:pr} shows the Period-Radius diagram produced in a similar way as the M-R relation Fig.~\ref{fig:mr}. Even in this case, TOI-179 b is located at the edge of the typical distribution of super-Earths/Neptune-like planets. Unlike the case of very young planets (age $\lesssim$ 50 Myr, depicted as red dots), for which we expect a radius evolution with time leading to a shrinking of the planetary structure (see e.g. the discussion in  \citealt{benatti2021}), TOI-179 b will probably remain in its current location of the diagram. In fact, as presented in Sect.~\ref{subs:evaporation}, we do not foresee a significant loss of the planetary atmosphere, therefore a reduction of the radius, driven by the high-energy irradiation from the star, because of its high-density.

\subsection{Planet evaporation} \label{subs:evaporation}
 
 In order to study the hydrodynamic stability of TOI-179 b, we applied our model described in e.g. \citet{benatti2021} and \citet{Maggio22}. In our model, we describe the time  evolution of planetary atmospheres of Neptunian planets using  the hydrodynamic-based  approach described in \citet{Kuby18} and adopting the evolution of the X-ray luminosity from \citet{penz2008}.  Assuming that all the planetary mass is concentrated  in the core, we found that  the core radius is 2.33 R$_{\oplus}$, consequently, using the formula carried out by \citet{Lopez14}, we estimated the  atmospheric mass fraction  obtaining f$_{atm}=0.173$\%. This small value of f$_{atm}$ suggests that the planet lost most of its primordial atmosphere. From the calculation of the Jeans parameter escape (e.g. \citealt{Fossati2017}) it results that the planet is stable against hydrodynamic evaporation, basically due to his large density.  
 
 We also investigated the past history of the planet seeking for what  mass and radius at 10 Myr a planet ends its evolution at the current stellar age with the observed mass and radius. We did not fix the core mass, but we explored a range of core masses and at each of them we assigned different values of the atmospheric mass fraction calculating consequently the initial planetary masses and radii. We have not been able to find a planet that ends its evolution with the mass and radius observed, other than the planet with the same mass and the same core mass previously estimated, but with a radius of 2.9 R$_{\oplus}$ that contracts down to the observed value only under the effect of the gravitational shrinking. This result is due to the fact that the most massive planets, that have a larger radius, even if they evaporate, cannot lose enough mass to reach the current observed value, while the less massive planets, due to the small corresponding radius,  tend to rapidly become hydrodynamically stable and therefore never reach the current mass (and radius) value.

\section{Summary and perspectives}
\label{sec:summary} 

We presented the validation and characterization of the TOI-179 system.
The central star is an active K2 star, with an age of 400$\pm$100 Myr determined by
a variety of indicators ans a solar chemical composition.
The star is orbited by a transiting planet with period 4.137436 days, mass 24$\pm$7 $M_{\oplus}$, radius 2.62$^{+0.15}_{-0.12} R_{\oplus}$,
and significant eccentricity (0.34$^{+0.07}_{-0.09}$), identified as a planet candidate by TESS and confirmed by extended RV time series obtained with HARPS. 
AO observations with SPHERE identified a low-mass companion at the boundary between
brown dwarfs and very low-mass stars (mass from luminosity 83$^{+4}_{-6} M_{J}$) at a very small projected separation (84.5  mas, 3.3 au at the distance of the star).
Coupling the imaging detection with the long-term RV trend and the astrometric signature
(proper motion anomaly considering Hipparcos and Gaia data), we constrained the orbit
of the low-mass companion, identifying two families of possible orbital solutions,
one with shorter period ($\sim$ 20-40 yr) and very high eccentricity ($\sim$0.9) and one with longer period ($\sim$ 100-120 yr) and moderate eccentricity ($\sim$0.3-0.5).
The closest family of orbital solutions would imply strong dynamical interactions with the transiting planet.
The architecture of the system is completed by a close pair (separation about 1") of K dwarfs at a projected separation of 3400 au from TOI-179.

The system is amenable for several follow-up observations.
First of all, the incomplete orbit coverage of the outer low-mass companion calls for
further observations both in imaging and radial velocity to firmly determine the orbital
solution and the dynamical mass of the object. Future Gaia releases will also provide
relevant contributions.
In the forthcoming years, with a larger projected separation from the central star,
spectroscopic observations of the companion will also become feasible, increasing the 
role of HD18599 B as a benchmark low-mass object.

The determination of the orbit of the outer companion will allow to understand its influence on the orbital parameters of the transiting planet, mandatory to understand the
origin of its eccentricity. The dynamical link between the two objects can also be investigated through Transit Time Variations (TTVs).  

The measure of the planetary orbital obliquity with respect to the stellar spin axis could be relevant to place further constraints on the system architecture and its evolution with time. This issue can be investigated through the measurement of the Rossiter-McLaughlin (RM) effect: the expected RM amplitude is $\sim 3.6$ \ms (by using Eq. 40 in \citealt{winn2010}), potentially detectable with ESPRESSO at VLT.

Further spectroscopic and photometric observations will also allow us to
reduce the uncertainties on planet parameters and then to better infer its evolutionary path (e.g. evaporation). Finally, the planet is also a possible target for atmospheric characterization, although its compact structure and high density  make it not particularly
highly ranked for this aim among planets of similar radii
\citep[a value of 52 is derived for the metric of transmission spectrum proposed by][]{kempton2018}.
The possibility of the occurrence of intense volcanic activity is of special interest in this context.
In brief, the system of TOI-179, presented in this paper, is potential high-merit laboratory for our understanding of the physical evolution of planets and other low-mass objects and of how the planet properties are influenced by dynamical effects and interactions with the parent star.

\begin{acknowledgements}
 
We acknowledge the use of public TESS Alert data from pipelines at the TESS Science Office and at the TESS Science Processing Operations Center.
Funding for the TESS mission is provided by NASA’s Science Mission directorate.
This work has made use of data from the European Space Agency (ESA) mission {\it Gaia} (\url{https://www.cosmos.esa.int/gaia}), processed by the {\it Gaia} Data Processing and Analysis Consortium (DPAC, \url{https://www.cosmos.esa.int/web/gaia/dpac/consortium}). Funding
for the DPAC has been provided by national institutions, in particular
the institutions participating in the {\it Gaia} Multilateral Agreement.
This work has made use of the SPHERE Data Centre, jointly operated by OSUG/IPAG (Grenoble), PYTHEAS/LAM/CeSAM (Marseille), OCA/Lagrange (Nice), Observatoire de Paris/LESIA (Paris), and Observatoire de Lyon/CRAL, and supported by a grant from Labex OSUG@2020 (Investissements d’avenir – ANR10 LABX56).
We thank F. Bouchy and X. Dumusque for managing the sharing of observing time between several programmes approved for HARPS in P103 and the observers of these programmes for carrying out observations of our target in their nights. 
This work has been supported by the PRIN-INAF 2019 "Planetary systems at young ages (PLATEA)" and ASI-INAF agreement n.2018-16-HH.0.  
DL acknowledges the support from ASI-INAF agreement n.2021-5-HH.0.
DN acknowledges the support from the French Centre National d'Etudes Spatiales (CNES). 
A.S. acknowledges support from the Italian Space Agency (ASI) under contract 2018-24-HH.0 "The Italian participation to the Gaia Data Processing and Analysis Consortium (DPAC)" in collaboration with the Italian National Institute of Astrophysics.
The authors became aware of a parallel effort on the characterization of TOI-179 by Vines et al. in the late stages of the manuscript preparations. Only submissions to arxiv were coordinated, and no analyses or results were shared prior to the acceptance of the papers.

\end{acknowledgements}

\bibliographystyle{aa} 
\bibliography{toi179} 
%

\appendix

\section{Stellar parameters}
\label{app:star}

We present in this Appendix the details of the determination of the stellar parameters for TOI-179 and its wide companions CD-56 593 A and B, briefly summarized in Sect \ref{sec:star} in main body of the paper.

\subsection{Spectroscopic analysis}
\label{sec:specanalysis}

We exploited the HARPS spectra described in Sect. \ref{sec:harps} for the stellar characterization.
A co-added spectrum was created;  
the composite spectrum has S/N per pixel at 6700~\AA~ of 180.

Our aim is to derive spectroscopic parameters and metallicity ([Fe/H]) via the equivalent width (EW) method and the standard iron line analysis. EWs were measured using $ARES$ \citep{sousa2007}, with frequent visual inspection and manual re-adjustment using the $IRAF$ task $splot$ (performing Gaussian fit). 
We adopted the linelist by \cite{dorazi2020}, which includes 80 Fe~{\sc i} and 17 Fe~{\sc ii} lines (see that paper for details on atomic parameters). We used the Castelli \& Kurucz grid of model atmosphere, with new opacities and no overshooting ({\sc odfnew}, \citealt{castelli2003}). Stellar parameters ($T_{\rm eff}$, $\log g$, microturbulence velocity V$t$) and metallicity were derived in a semi-automated way using the {\sc moog} \citep{sneden73} python wrapper qoyllur-quipu ($q^2$) code developed by I. Ram\'irez\footnote{The code and the tutorial are made available through \url{ https://github.com/astroChasqui/q2.}}.
We first derived parameters and metallicity for the HARPS-N solar spectrum obtained 
by \citet{biazzo2022pub}  and found $T_{\rm eff}$=5790$\pm$43 K, $\log g_\odot$=4.44$\pm$0.10 dex, V$t$=0.97$\pm$0.07 km s$^{-1}$, $\log$ n(Fe {\sc i})= 7.49$\pm$0.03 dex.
The internal uncertainties reported on atmospheric parameters come  from errors on the slopes for $T_{\rm eff}$ and V$t$, whereas from the ionisation balance condition of iron for surface gravity, as also routinely done in our previous works (\citealt{biazzo2015}; \citealt{dorazi2020}, and references therein).

We found $T_{\rm eff}$=5172$\pm$60 K, $\log$g=4.54$\pm$0.09 dex, V$t$=1.09$\pm$0.14 kms$^{-1}$, and [Fe/H]=0.00$\pm$0.08 dex (total error). 
Previously, \citet{mortier2013} reported
$T_{\rm eff}$=5154 K (from B-V color), and [Fe/H]=-0.01 dex, (from their calibration of CCF parameters) while \citet{jenkins2008} derived [Fe/H]=-0.01 dex from their analysis of FEROS spectra ($T_{\rm eff}$ not listed).
Comparison with photometric sequences by \citet{pecaut2013} \footnote{Updated version available at \url{https://www.pas.rochester.edu/~emamajek/EEM_dwarf_UBVIJHK_colors_Teff.txt}, hereafter referred as Mamajek tables} yields a photometric $T_{\rm eff}$=5118$\pm$50 K, assuming no reddening as the target is at only 38 pc from the Sun.
Independent measurements then agree fairly well.
In the following, we adopt the mean of our spectroscopic and photometric determinations, $T_{\rm eff}$=5145 K, with a conservative error of 50 K.

\subsection{Chemical abundances} \label{sec:chem}

We have determined elemental abundances for proton-capture, $\alpha$ and iron-peak elements Na~{\sc i}, Mg~{\sc i}, Al~{\sc i}, Ca~{\sc i}, Ti~{\sc i}, Ti~{\sc ii}, and Ni~{\sc i} by using the linelist of \cite{dorazi2020} and the same code and model atmosphere described in Section \ref{sec:specanalysis}. As previously done for the [Fe/H] determination, we first analysed the solar spectrum and obtained the elemental abundances as reported in Table\,\ref{tab:elements}. 
For Na {\sc i} lines 6154/6160 \AA~ we applied Non-Local Termodynamic Equlibrium (NLTE) corrections from the INSPECT database (\citealt{lind2011}).
We calculated internal errors due to EW measurements (the first error in Table\,\ref{tab:elements}, which is the error on the mean from different lines) and the uncertainty related to atmospheric parameters. 
All the abundances exhibit a solar-scaled pattern, with only a marginal enhancement in calcium ([Ca/H]=0.07 dex) and titanium ([Ti/H]=0.07 from the average of neutral and singly-ionised atomic lines). This is probably related to blending effects of the lines and/or continuum displacements at lower temperatures.

\begin{table}[htbp]
    \centering
    \begin{tabular}{lccr}
    \hline\hline
    Species & N-lines & Sun & TOI-179 \\
    \hline
    Na~{\sc i}$_{\rm NLTE}$ &  2 & 6.21$\pm$0.01$\pm$0.01 & 6.21$\pm$0.01$\pm$0.02 \\
    Mg~{\sc i}              &  2 & 7.56$\pm$0.03$\pm$0.01 & 7.59$\pm$0.03$\pm$0.02 \\ 
    Al~{\sc i}              &  2 & 6.51$\pm$0.01$\pm$0.01 & 6.56$\pm$0.01$\pm$0.01 \\
    Si~{\sc i}              & 11 & 7.49$\pm$0.02$\pm$0.02 & 7.47$\pm$0.01$\pm$0.02 \\
    Ca~{\sc i}              &  9 & 6.32$\pm$0.04$\pm$0.01 & 6.39$\pm$0.03$\pm$0.01 \\
    Ti~{\sc i}              & 52 & 4.96$\pm$0.04$\pm$0.02 & 5.05$\pm$0.04$\pm$0.03\\
    Ti~{\sc ii}             & 25 & 4.97$\pm$0.05$\pm$0.02 & 5.03$\pm$0.06$\pm$0.04\\
    Ni~{\sc i}              & 16 & 6.24$\pm$0.01$\pm$0.02 & 6.24$\pm$0.02$\pm$0.03\\
    \hline\hline
    \end{tabular}
    \caption{Elemental abundances (with corresponging number of lines) as derived for the solar spectrum and TOI-179.}\label{tab:elements}
    \label{tab:my_label}
\end{table}

\subsection{Lithium} \label{sec:lithium}

\begin{figure}[htbp]
\includegraphics[width=0.9\columnwidth]{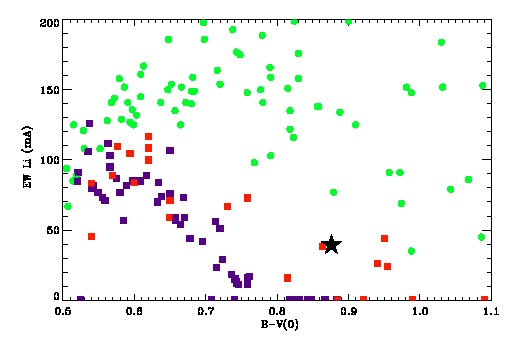}
\caption{\label{fig:ewli} 
EW Li of TOI-179 (black star), compared to the members of the Pleiades (green circles), Hyades (purple squares), and Ursa Major (red squares).}
\end{figure}

We also exploited the HARPS spectrum to measure the equivalent width (EW) of the Li 6708\AA\, doublet, that results 39.3$\pm$4.5 m\AA.
The Li EW of TOI-179 is intermediate between those of Pleiades and Hyades of similar colors (Fig.~\ref{fig:ewli}), close but slightly smaller than that of M34 members \citep{jones1997,gondoin2014}, and close to the upper envelope of the members of Ursa Major moving group \citep{soderblom1993,ammler2009}.

\begin{figure}[htbp]
\includegraphics[width=0.9\columnwidth]{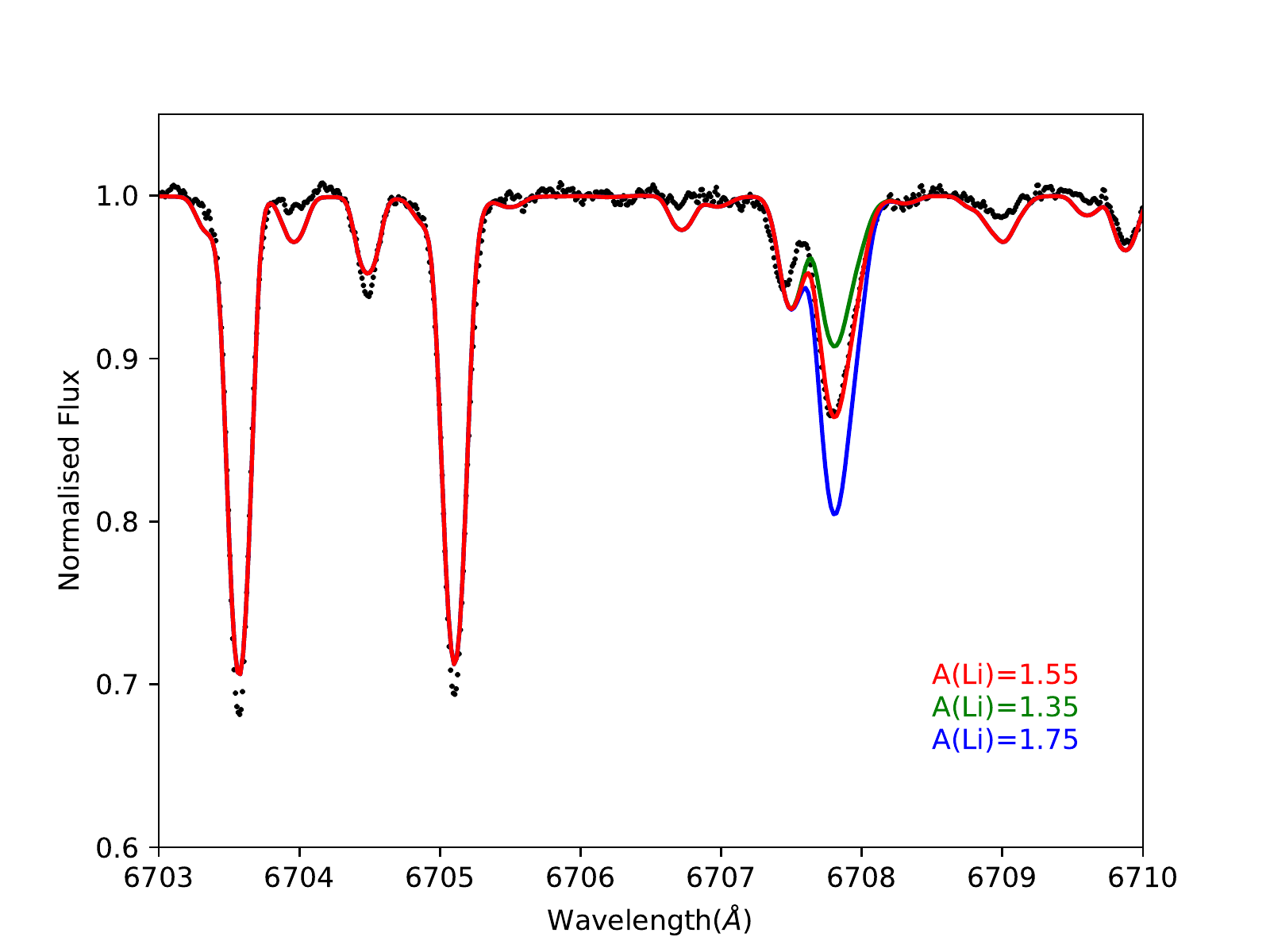}
\caption{Comparison between observed (black dots) and synthetic (green, red, and blue lines) spectra around the Li~{\sc i} line at 6708 \AA.}
\label{fig:lithium_synth}
\end{figure}

We employed the stellar parameters derived in Section \ref{sec:specanalysis} and performed spectral synthesis calculations in order to derive the Li abundance. We used the driver $synth$ in MOOG (2019 version) and the line list around the Li doublet at 6708\AA\ as done in our previous investigations. 
By adopting a FWHM of 0.06 \AA\ for the instrumental profile, limb darkening of 0.65, we have determined 
A(Li~{\sc i})=1.55$\pm$0.08 dex (see Fig. \ref{fig:lithium_synth}).

\subsection{Projected rotational velocity} \label{sec:vsini}

We derived the projected rotational velocity $v \sin i$ exploiting the HARPS spectra with
two independent methods.
From the spectral synthesis in the region of the Li doublet, we derived $v \sin i$ = 4.5 $\pm$ 0.5 km s$^{-1}$.
From a preliminary calibration of the observed FWHM of the CCF into rotational velocity
for HARPS spectra, we obtained $v \sin i$ = 4.50 $\pm$ 0.41 km s$^{-1}$.
Determinations available in the literature are 4.3 km s$^{-1}$ \citep{jenkins2011} 
and 4.6 km s$^{-1}$ \citep{grandjean2021}, in good agreement with our determinations.

\subsection{Chromospheric and coronal activity} 
\label{sec:activity}

\begin{figure}[htbp]
\includegraphics[width=\columnwidth]{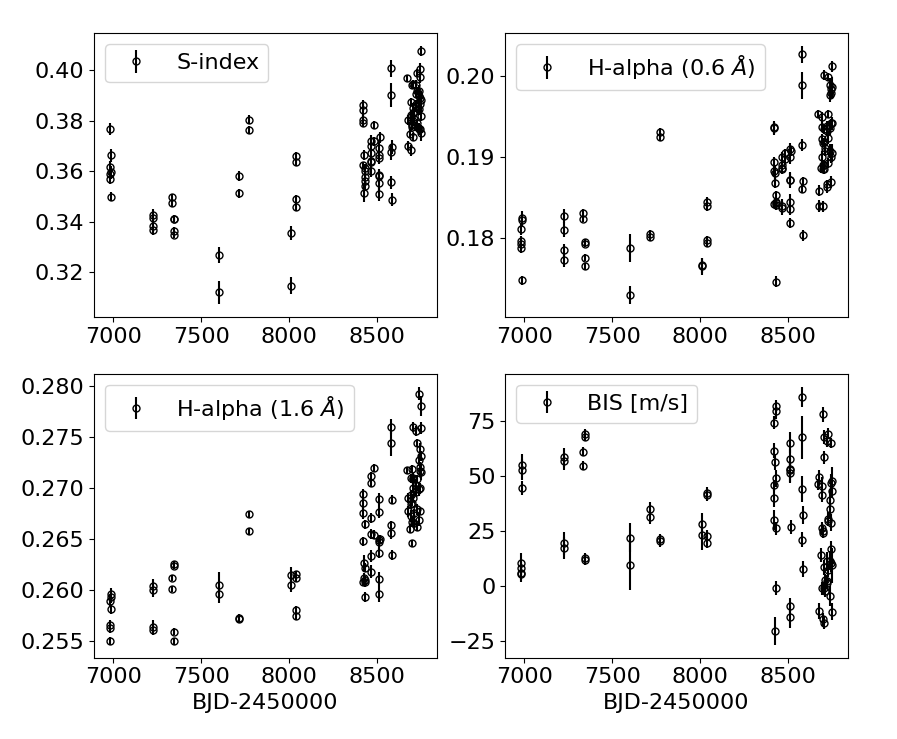}
\caption{\label{fig:activity} 
Time series of Ca II H\&K instrumental S index, H${\alpha}$, and bisector velocity span (BIS).}
\end{figure}

We measured the instrumental S index (Ca II H\&K) and H$\alpha$ index using the tool ACTIN (v1.3.9; \citealt{gomesdasilva18}). We calculated the H$\alpha$ index adopting both the definitions labeled as ``Ha06'' and ``Ha16'' in the line configuration file, denoting indices with 0.6 and 1.6~\AA~central bandwidths, respectively. 
The most relevant signal is a long-term modulation, likely ascribed to an activity cycle, which is nearly linear in H$\alpha$ (especially for the case of the Ha16 index), while a possible turnover is observed for the S index (Fig. \ref{fig:activity}). Such a difference may be real, and linked to the different kinds of active regions to which each indicator is sensitive, but might also be due to the modification of the HARPS set-up in 2015 \citep{locurto2015}.
A nearly linear upward trend is also seen in the RV time series (see Sect.~\ref{sec:rv_tess_fit_toi179}). Dedicated tests were performed and presented in Sect.~\ref{sec:trend} supporting the Keplerian origin of the RV trend. 

After removing the long-term trend, a significant periodicity at 8.64 d appears on the residuals of the S index timeseries, corresponding to the rotation period of the star, while no 
significant short-term periodicity is found for H$\alpha$.

As the ACTIN S index is not calibrated into the standard M. Wilson scale \citep{baliunas1995},
we also obtained calibrated $S_{MW}$ index using the procedure adopted in \citet{desidera2015}. The mean values result $S_{MW}$=0.607 and, with B-V color
from Table \ref{t:star_param} and \citet{noyes1984} transformation, $\log R'_{HK}$=-4.35.
The minimum and maximum values considering the prominent long-term activity cycle are
$\log R'_{HK}$=-4.43 and -4.30, respectively. The sparse measurements by \citet{gray2006,jenkins2008,jenkins2011} are within this distribution.
The activity level results intermediate between Hyades and Pleiades/AB Dor stars of similar colors.

From the CHANDRA observations described in Sect.~\ref{sec:chandra}, we obtained for TOI-179 a kT of 0.56 keV and
an unabsorbed flux of $1.79\times10^{-13}$ erg s$^{-1}$ cm$^{-2}$ in the ROSAT band 0.1--2.4 keV which
corresponds to a X-ray luminosity of $L_X \approx 3.23\times10^{28}$ erg s$^{-1}$  at a distance of 38.6 pc.  
The normalization constant of the best fit model corresponds to an emission measure 
of $2.7\times10^{51}$ cm$^{-3}$.
Coupled with the stellar luminosity from Table \ref{t:star_param}, this implies 
$R_X = \log (L_X/L_{bol}) = -4.65$, slighty above the median value of Hyades members of similar colors.

\subsection{Photometric variability and rotation period} \label{sec:rot}

\begin{figure}[htbp]
\includegraphics[width=8.7cm,angle=180]{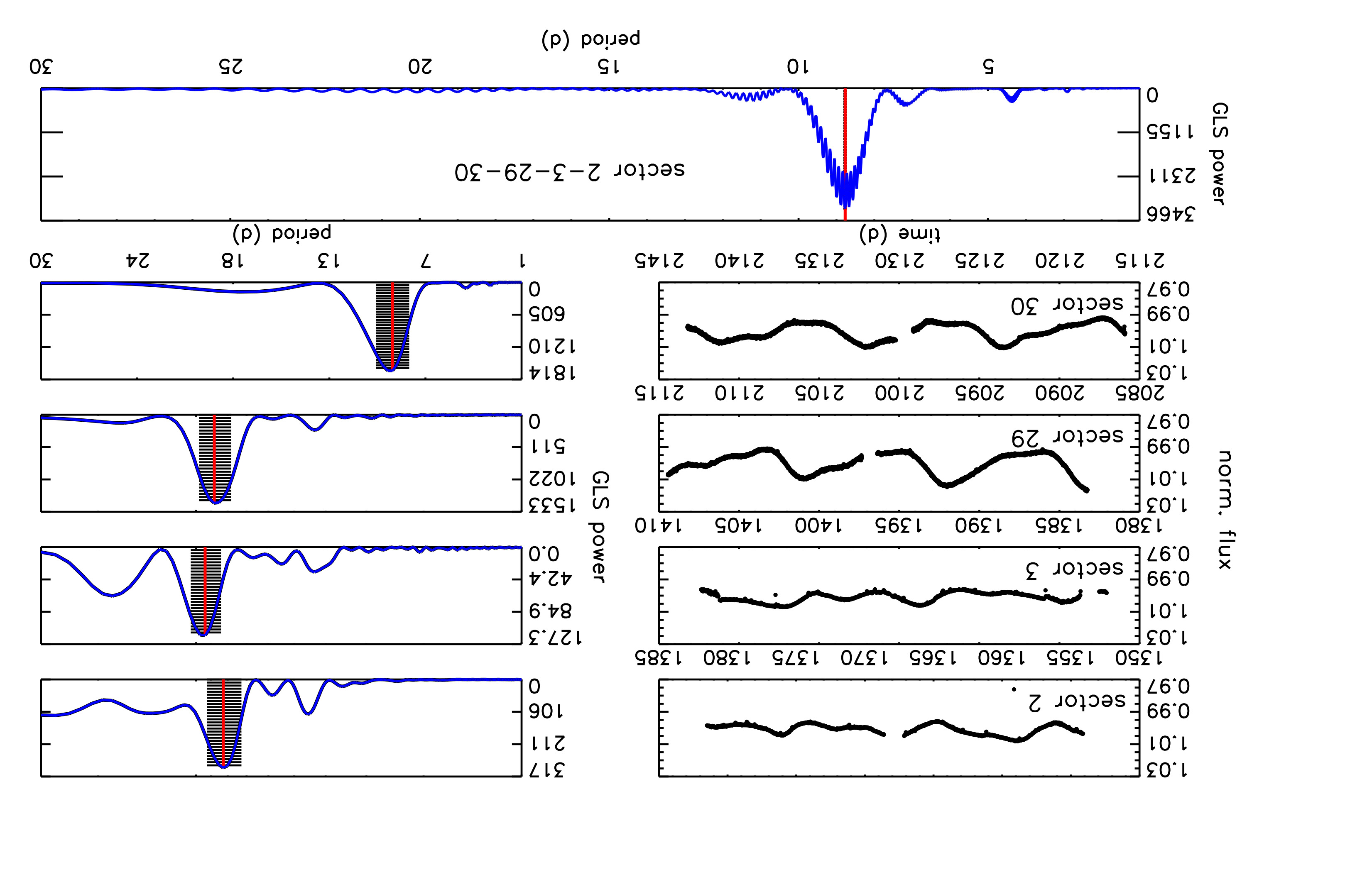}
\caption{\label{fig:TESS_GLS} 
TESS lightcurves in Sectors 2, 3, 29, and 30 (from top to bottom) represented as normalized flux versus time (left column), and corresponding GLS periodograms (right column) with indication of the measured periodicity (vertical red line) and its uncertainty (shaded region). Similar periodograms are obtained when using CLEAN (see \citet{messina2022}). The bottom plot shows the periodogram when all four Sectors are combined into a unique timeseries. }
\end{figure}

Three determinations of rotation period of TOI-179 are available in the literature. 
 \citet{oelkers2018} measured a period of 8.693d from KELT photometric data, \citet{howard2021}
 obtained P=8.64$\pm$0.04 d from their analysis of TESS and Evryscope data, while \citet{canto2020}
 derived P=8.49$\pm$0.86 d considering TESS data only (sectors 2-3).
Our analysis of TESS light curve (four sectors considered independently), performed with the methods described in
\citet{messina2022} yields a photometric period of 8.84$\pm$0.29\,d (see Fig.\,\ref{fig:TESS_GLS}), in general agreement with the literature results. 
The analysis of the time series of HARPS RV and the corresponding activity indicators
also provides additional estimates of rotation periods.
The Gaussian process regression analysis  of the RV time series described in Sect. \ref{sec:rv_tess_fit_toi179} yields a period of 8.72$\pm$0.04 d,  while
the S index periodogram analysis (Sect. \ref{sec:activity}) indicates a period of 8.64 d.
Averaging the above determinations, we derive P= 8.73$\pm$0.07 d.

\begin{figure*}[htbp]
\centering
\includegraphics[width=14cm,angle=180]{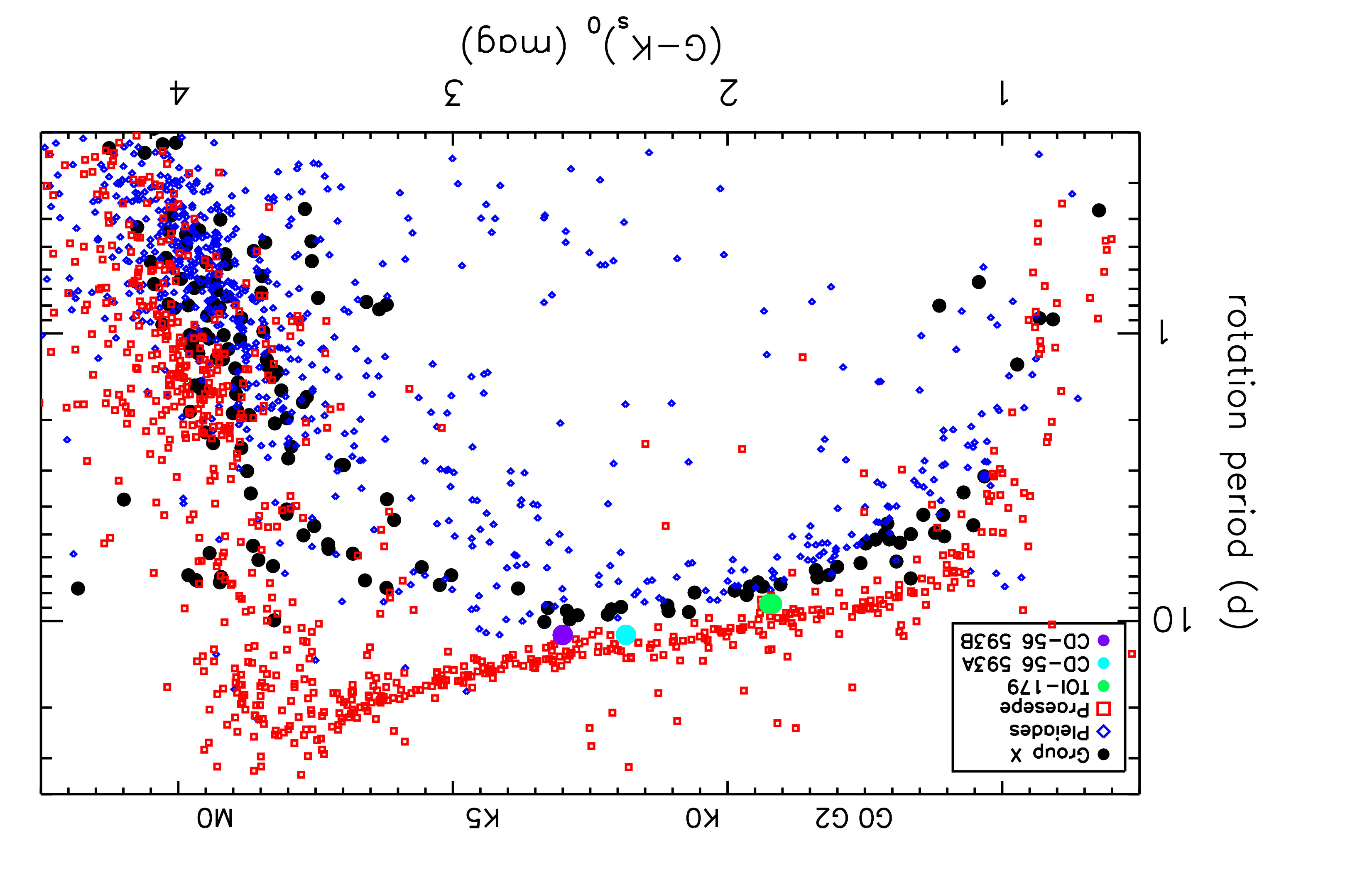}
\caption{\label{fig:prot} 
Rotation period of TOI-179 and CD-56 593A and B vs (G$-$K)$_0$ color (green, cyan, and purple filled circles, respectively), compared to the members
of the Pleiades and Praesepe open clusters (\citealt{Rebull16}, \citeyear{Rebull17}, plotted as 
blue diamonds and red squares, respectively) and Group X moving group \citep[][ plotted as black circles]{messina2022}. Only one photometric period is measured for CD-56 593, spatially unresolved in TESS images, and we are unable to conclude to which component it belongs to. The data are plotted in the figure for the colors of both components. }
\end{figure*}

The adopted rotation period results slightly longer than the rotation-color sequence
of the 300 Myr Group X recently obtained by \citet{messina2022}. It is clearly distinct from the Pleiades sequence on the young side and from Hyades and Praesepe on the old side (Fig. \ref{fig:prot}). From the \citet{mamajek2008} calibration, 
the resulting age is 390 Myr.

Finally, to further constrain the long-term activity variations we also checked the ASAS light curve \citep{Pojmanski97}, which covers about 9 years between 2000 and 2009. A small downward trend of 0.003$\pm$0.0003 mag/yr is retrieved, without indications of large long-term photometric variations and significant periodicities due to an activity cycle.

\subsection{Kinematics and membership to groups} 
\label{sec:kin}

TOI-179 is not a member of known moving groups according to our analysis using 
the BANYAN $\Sigma$ code by \citet{banyansigma}
\footnote{
\url{http://www.exoplanetes.umontreal.ca/banyan/banyansigma.php}}.
Nevertheless, its kinematic parameters are well within the kinematic box of young stars \citep{montes2001}. A full-sky search in the \citet{smart2021} catalog for stars with similar space velocity components U,V,W with parallax larger than 15 mas yields only 13 stars with difference in each of the space velocities of less than 2 km/s with respect to TOI-179\footnote{The wide companion CD-56 593 (see Sect. \ref{sec:bin}) is not included in this output because it lacks radial velocity determinations}. These stars are scattered over the whole sky; only some of them have indication of being moderately active and young, but likely older than TOI-179, the majority has no indication of youth beside kinematics and one is clearly older being a red giant. We conclude that our target is not part of any group or association.

\subsection{Multiplicity: the wide companions CD-56 593 A and B} 
\label{sec:bin}

TOI-179 is a known triple system. CD-56 593 is a close pair (1.50" = 58 au projected separation at the date of the first observation listed in WDS \citep{wds}, epoch 1911; 1.20"= 46 au projected separation in 1991 \citep{tychobin}; $\Delta V$ = 0.76 mag) at 87.5" almost exactly south ($PA=182$~degree) from TOI-179, corresponding to a physical projected separation of 3380 au. The close binary is not resolved in Gaia DR2 while there are separate entries in Gaia EDR3, with a projected separation of 0.90", a significant position angle shift (194 deg, compared to 185 deg in 1911), and a magnitude difference $\Delta G$=0.55 mag. The differences in the proper motion between the components and with respect to TOI-179 are linked to orbital motion and they are small enough to firmly conclude on the physical association to TOI-179. The common proper motion was previously noted by \cite{tokovinin2012} and \citet{mugrauer2020}. No other comoving objects are found in Gaia EDR3 within 1 deg. The presence of close companions will be discussed in Sect.~\ref{sec:bd}.

The position of the components on the $M_G$ vs $BP-RP$ color-magnitude diagram agrees within errors with the standard main sequence in Mamajek tables. The expected spectral types are K5 and K6/K7 (closer to K7) for CD-56 593A and CD-56 593B, respectively, and the stellar masses  0.70 and 0.66 M$_\odot$\ for CD-56 593A and CD-56 593B, respectively. The system is then formed by three K dwarfs.

CD-56 593 has a photometric rotation period (which most likely belongs to the brighter component CD-56 593A) of 11.20 d \citep{oelkers2018}. We also analyzed the TESS data as done for TOI-179 in Sect.~\ref{sec:rot}. We found a period similar to the literature one for Sector 2 data, while a period close to a half of this value was derived for Sector 30 data, most likely an harmonic of the true period linked to the distribution of starspots on stellar surface. The period in Sector 29 is instead uncertain as it appears dominated by the evolution of active regions or variable contributions from the two components. Data in Sector 3 are of poor quality.
Combining the data, we adopt a period of 11.2$\pm$0.9\,d.
After deblending of the Ks magnitude following Mamajek tables, this period results slightly slower than the sequence of the 300 Myr old Group X, as found above for TOI-179, independently of the component to which the observed periodicity belongs to
(Fig. \ref{fig:prot}).

CD-56 583 is also detected in the CHANDRA images presented in Sect.~\ref{sec:chandra}, with spatially unresolved emission between the two components. We obtained a temperature of 0.67 keV and an unabsorbed flux of 1.13 $\times$ 10$^{-13}$~erg s$^{-1}$ cm$^{-2}$ in the ROSAT band, corresponding to a system X-ray luminosity of 2.03 $\times$ 10$^{28}$ erg/s. We also obtained a value of emission measure for each component equal to $\sim8\times10^{50}$ cm$^{-3}$. An estimated value of the X-ray luminosities of the individual components, assuming they have the same $R_X$ ratio ($R_X$=-4.74, from the individual luminosities of 0.181 and 0.112 L$_{\odot}$, for CD-56 593A and CD-56 593B, respectively, derived as in Sect.~\ref{a:massradius}), is 1.25 $\times$ 10$^{28}$ erg/s and 0.78 $\times$ 10$^{28}$ erg/s for CD-56 593A and CD-56 593B, respectively. These values are similar to Hyades members of similar spectral type.

\subsection{Age} \label{a:age}

The age indicators of TOI-179 consistently support an age older than the Pleiades and younger than the Hyades. Comparison with the 300 Myr old Group X moving group indicates a slighty older age. Isochrone fitting supports an age close to the main sequence, consistently with the indirect age indicators, both for TOI-179 and for CD-56 593 A and B (within the large color uncertanties for these latter objects).
Considering these results, we rely on indirect indicators, of which the most sensitive given the age and spectral type of the object are lithium and rotation period. We adopt an age of 400 Myr with limits between 300 to 500 Myr. The results of the wide binary companion CD-56 593 (CMD position, rotation period, and X-ray emission) are also consistent with this age estimate.

\subsection{Mass, radius, and system orientation} \label{a:massradius}

We exploited the PARAM web interface \citep{dasilva2006}\footnote{ \url{http://stev.oapd.inaf.it/cgi-bin/param_1.3} } to infer the stellar mass. As in \citet{desidera2015}, we restricted the values of stellar ages to those allowed by indirect methods. This shifts the stellar mass upward by about 0.03~M$_\odot$. The final adopted mass results 0.863$\pm$0.020~M$_\odot$.

The stellar radius was derived through Stefan-Botzmann law, as done in \citet{carleo2021}, using $T_{\rm eff}$ from Sect.~\ref{sec:specanalysis}, V magnitude from Table \ref{t:star_param}, and bolometric correction from Mamajek tables. The adopted radius results 0.767$\pm$0.024~R$_\odot$, slightly smaller but in agreement to better than one $\sigma$ with the results listed in Gaia DR2, PIC and TIC catalogs
\citep{pic,tic}, and EXOFOP database\footnote{\url{https://exofop.ipac.caltech.edu/tess/target.php?id=207141131}}.

Coupling our adopted radius with the rotation period and projected rotational velocity from Sect. \ref{sec:vsini} yields a stellar inclination of $90^{+0}_{-30} $ deg, with
the expected equatorial velocity (4.44$\pm$0.15 km/s) nearly identical to the observed $v \sin i$ (4.5$\pm$0.5 km/s). Therefore, the star appears to be seen close to edge-on. Alignment with the orbit of the transiting planets ($i_b$=87.5 deg, see Table \ref{t:toi-179_pl_param}) is then likely, although  a moderate misalignment is possible within the errorbars.

\section{Radial velocity and activity indicators time series}
\label{a:table}

\longtab[1]{
\begin{longtable}{lccccccc}
\caption{\label{tab:rv} Radial velocities from TOI-179 HARPS spectra
with uncertainties as obtained from the TERRA pipeline. The corresponding measure of the bisector span is provided by the HARPS DRS. The H$\alpha$ and the S-index with corresponding uncertainties are obtained with ACTIN.}\\
\hline\hline

\hline
\endfirsthead
\caption{continued.}\\
\hline\hline
 BJD$_{\rm TDB}$ - 2450000    & RV   &    err$_{\rm RV}$     &   BIS  &  H$\alpha$[1.6 \AA] & err$_{H\alpha [1.6 \AA]}$ & S-index & err$_{\rm S-index}$ \\
                              & [m s$^{-1}$]    &  [m s$^{-1}$]   &  [km s$^{-1}$]  &  - & - & - & - \\
\hline
\endhead
\hline
\endfoot
6980.69590864 & 2.0808 & 0.9411 & 0.005250 & 0.2566 & 0.0004 & 0.3663 & 0.0021 \\
6980.70195011 & 0.0000 & 0.8142 & 0.005890 & 0.2550 & 0.0004 & 0.3691 & 0.0028 \\
6982.67878363 & 3.2470 & 1.1515 & 0.008210 & 0.2563 & 0.0005 & 0.3581 & 0.0027 \\
6982.68488297 & 4.1855 & 1.2608 & 0.010610 & 0.2590 & 0.0006 & 0.3653 & 0.0025 \\
6987.63091820 & -29.3223 & 1.3290 & 0.055070 & 0.2593 & 0.0006 & 0.3553 & 0.0031 \\
6987.63707540 & -30.5069 & 1.4453 & 0.052730 & 0.2596 & 0.0006 & 0.3584 & 0.0021 \\
6988.69275587 & -32.0267 & 0.8648 & 0.044450 & 0.2581 & 0.0004 & 0.3511 & 0.0029 \\
7223.88142723 & -43.2483 & 2.0742 & 0.017150 & 0.2600 & 0.0007 & 0.3734 & 0.0021 \\
7223.89251559 & -38.0497 & 2.2566 & 0.019390 & 0.2604 & 0.0007 & 0.3426 & 0.0025 \\
7224.88157860 & -71.0729 & 1.7280 & 0.058330 & 0.2564 & 0.0006 & 0.3383 & 0.0026 \\
7224.88963445 & -69.1581 & 1.4157 & 0.056530 & 0.2561 & 0.0006 & 0.3367 & 0.0021 \\
7333.76625907 & -78.6104 & 0.9334 & 0.054490 & 0.2611 & 0.0003 & 0.3413 & 0.0019 \\
7333.77693008 & -78.3213 & 1.0476 & 0.060910 & 0.2601 & 0.0003 & 0.3473 & 0.0013 \\
7342.74185760 & -98.4356 & 1.1666 & 0.068750 & 0.2623 & 0.0003 & 0.3497 & 0.0014 \\
7342.75252857 & -96.7108 & 1.1087 & 0.067700 & 0.2625 & 0.0003 & 0.3411 & 0.0014 \\
7344.69134515 & -27.4767 & 1.0305 & 0.012730 & 0.2550 & 0.0003 & 0.3412 & 0.0013 \\
7344.70191194 & -28.9744 & 1.2010 & 0.011800 & 0.2559 & 0.0003 & 0.3364 & 0.0014 \\
7600.82445252 & -15.0062 & 2.7856 & 0.021910 & 0.2596 & 0.0008 & 0.3348 & 0.0014 \\
7600.83648994 & -17.2872 & 4.7693 & 0.009610 & 0.2605 & 0.0013 & 0.3269 & 0.0032 \\
7713.80082617 & -26.4906 & 1.5446 & 0.031270 & 0.2572 & 0.0004 & 0.3120 & 0.0046 \\
7713.81152026 & -25.8336 & 1.1222 & 0.035130 & 0.2573 & 0.0004 & 0.3580 & 0.0020 \\
7770.61269041 & 0.4020 & 1.1772 & 0.021300 & 0.2674 & 0.0003 & 0.3512 & 0.0019 \\
7770.62324557 & 2.3310 & 1.3228 & 0.020580 & 0.2658 & 0.0003 & 0.3763 & 0.0016 \\
8007.71941096 & -35.0594 & 1.9859 & 0.027950 & 0.2605 & 0.0007 & 0.3803 & 0.0018 \\
8007.73070737 & -37.9881 & 2.4466 & 0.023290 & 0.2614 & 0.0008 & 0.3356 & 0.0029 \\
8035.62399563 & -12.9255 & 1.1087 & 0.022640 & 0.2575 & 0.0004 & 0.3148 & 0.0034 \\
8035.63478257 & -13.3406 & 0.9869 & 0.019700 & 0.2580 & 0.0004 & 0.3460 & 0.0016 \\
8039.62904707 & -13.0419 & 1.1157 & 0.041150 & 0.2612 & 0.0004 & 0.3490 & 0.0015 \\
8039.63992657 & -12.7141 & 1.1839 & 0.042290 & 0.2615 & 0.0004 & 0.3660 & 0.0016 \\
8417.56050911 & 15.8051 & 1.6301 & 0.039980 & 0.2675 & 0.0006 & 0.3637 & 0.0016 \\
8417.76713698 & 1.5168 & 1.6723 & 0.030110 & 0.2685 & 0.0006 & 0.3968 & 0.0014 \\
8418.55583647 & -36.2440 & 1.3023 & 0.061410 & 0.2694 & 0.0005 & 0.3701 & 0.0020 \\
8418.72027757 & -40.3748 & 1.4555 & 0.074110 & 0.2648 & 0.0004 & 0.3801 & 0.0019 \\
8419.68677382 & -18.8122 & 1.2185 & 0.045670 & 0.2608 & 0.0004 & 0.3749 & 0.0017 \\
8422.83131156 & 30.2399 & 2.3066 & -0.020340 & 0.2626 & 0.0008 & 0.3815 & 0.0017 \\
8423.74520254 & -8.5072 & 1.4759 & 0.056360 & 0.2612 & 0.0005 & 0.3821 & 0.0016 \\
8429.76226288 & -12.5505 & 1.5425 & 0.048860 & 0.2593 & 0.0005 & 0.3872 & 0.0016 \\
8430.74908821 & 2.3853 & 1.7303 & 0.026080 & 0.2664 & 0.0004 & 0.3682 & 0.0021 \\
8431.67791694 & 26.4825 & 1.4431 & -0.000970 & 0.2622 & 0.0004 & 0.3774 & 0.0013 \\
8432.68523840 & -24.2928 & 1.2659 & 0.081430 & 0.2608 & 0.0004 & 0.3943 & 0.0016 \\
8432.71563107 & -24.8233 & 1.3730 & 0.079500 & 0.2609 & 0.0005 & 0.3784 & 0.0020 \\
8464.72530016 & -9.8082 & 1.6310 & -         & 0.2633 & 0.0006 & 0.3786 & 0.0013 \\
8464.73352901 & -6.8396 & 1.7945 & -         & 0.2618 & 0.0006 & 0.3736 & 0.0017 \\
8465.69725272 & 35.3156 & 1.5869 & -         & 0.2711 & 0.0004 & 0.3797 & 0.0016 \\
8465.70501861 & 33.8672 & 1.4372 & -         & 0.2705 & 0.0004 & 0.3830 & 0.0016 \\
8466.68945895 & -26.6290 & 1.3620 & -        & 0.2670 & 0.0005 & 0.3944 & 0.0018 \\
8466.69655358 & -26.1766 & 1.6262 & -        & 0.2655 & 0.0005 & 0.3850 & 0.0021 \\
8481.60390690 & -0.1308 & 1.4002 & -         & 0.2654 & 0.0005 & 0.3833 & 0.0018 \\
8483.60647679 & 7.4831 & 1.1414 & -          & 0.2719 & 0.0004 & 0.3808 & 0.0016 \\
8509.53123346 & -34.2170 & 1.6320 & 0.053220 & 0.2649 & 0.0005 & 0.3907 & 0.0016 \\
8509.57920635 & -34.3604 & 2.0194 & 0.064890 & 0.2689 & 0.0006 & 0.3865 & 0.0015 \\
8509.59794413 & -39.5850 & 1.8169 & 0.051320 & 0.2676 & 0.0006 & 0.3859 & 0.0023 \\
8510.55122193 & -42.5098 & 1.9474 & 0.057400 & 0.2596 & 0.0007 & 0.3943 & 0.0016 \\
8510.58082740 & -37.6804 & 2.6694 & 0.052610 & 0.2611 & 0.0007 & 0.3990 & 0.0014 \\
8511.54866504 & 35.0932 & 1.5730 & -0.008900 & 0.2636 & 0.0005 & 0.3919 & 0.0014 \\
8511.60528350 & 38.0889 & 1.6384 & -0.014010 & 0.2647 & 0.0006 & 0.3790 & 0.0017 \\
8512.55084217 & 15.5591 & 1.4106 & 0.026710 & 0.2649 & 0.0004 & 0.3775 & 0.0017 \\
8576.48050990 & 0.4351 & 2.4353 & 0.044050 & 0.2760 & 0.0007 & 0.3766 & 0.0024 \\
8576.51237350 & -5.4469 & 3.4031 & 0.067490 & 0.2744 & 0.0012 & 0.3848 & 0.0025 \\
8579.49225408 & -44.4058 & 1.9851 & 0.085650 & 0.2655 & 0.0006 & 0.3899 & 0.0026 \\
8580.47997019 & 14.8276 & 1.4775 & 0.020970 & 0.2663 & 0.0004 & 0.3918 & 0.0024 \\
8581.48412227 & 5.2024 & 1.7351 & 0.032370 & 0.2634 & 0.0005 & 0.3973 & 0.0022 \\
8582.47372628 & 34.2823 & 1.6718 & 0.007800 & 0.2689 & 0.0004 & 0.3890 & 0.0023 \\
8670.93364852 & 6.9920 & 1.0773 & 0.046050 & 0.2717 & 0.0004 & 0.3866 & 0.0014 \\
8673.87429327 & 31.5917 & 1.4553 & -0.011120 & 0.2677 & 0.0006 & 0.3766 & 0.0019 \\
8674.94274813 & -9.4457 & 1.3615 & 0.049300 & 0.2690 & 0.0005 & 0.4005 & 0.0022 \\
8684.87453781 & 17.8999 & 1.2278 & 0.013930 & 0.2660 & 0.0005 & 0.3750 & 0.0032 \\
8689.86328901 & 2.8129 & 1.1143 & 0.041050 & 0.2684 & 0.0004 & 0.3819 & 0.0022 \\
8690.87555738 & 39.9991 & 1.2297 & -0.001130 & 0.2693 & 0.0005 & 0.3880 & 0.0039 \\
8692.81900246 & 14.0329 & 1.0153 & 0.026230 & 0.2709 & 0.0004 & 0.4076 & 0.0019 \\
8693.85287935 & -2.9615 & 1.5832 & 0.045490 & 0.2672 & 0.0006 & 0.3861 & 0.0021 \\
8694.84459014 & 0.2584 & 0.9592 & 0.024070 & 0.2686 & 0.0003 & 0.3842 & 0.0026 \\
8695.87172996 & 50.5443 & 1.3158 & -0.015140 & 0.2718 & 0.0004 & 0.3804 & 0.0017 \\
8697.78683736 & -35.8855 & 1.4054 & 0.077990 & 0.2667 & 0.0005 & 0.3793 & 0.0017 \\
8698.82266843 & 8.1072 & 1.0585 & 0.025040 & 0.2646 & 0.0003 & 0.3624 & 0.0013 \\
8699.88169423 & 37.6680 & 1.2428 & -0.016730 & 0.2710 & 0.0005 & 0.3513 & 0.0034 \\
8700.87854176 & 0.0000 & 1.1900 & 0.058290 & 0.2700 & 0.0004 & 0.3666 & 0.0019 \\
8701.80598865 & 26.7346 & 1.3374 & -0.000930 & 0.2690 & 0.0004 & 0.3542 & 0.0021 \\
8702.76752183 & -6.0488 & 1.2357 & 0.067500 & 0.2760 & 0.0005 & 0.3561 & 0.0018 \\
8704.79732613 & 38.6818 & 1.4503 & 0.008390 & 0.2709 & 0.0005 & 0.3612 & 0.0016 \\
8707.90966982 & 28.4307 & 1.3332 & -0.002070 & 0.2666 & 0.0005 & 0.3600 & 0.0015 \\
8708.76447295 & 40.6240 & 1.1993 & 0.002630 & 0.2675 & 0.0004 & 0.3577 & 0.0020 \\
8718.74961073 & 36.4584 & 1.2453 & 0.001510 & 0.2756 & 0.0005 & 0.3602 & 0.0024 \\
8719.82471466 & -22.4962 & 1.1639 & 0.065630 & 0.2696 & 0.0004 & 0.3639 & 0.0026 \\
8720.84621731 & 11.7789 & 1.6623 & 0.011170 & 0.2703 & 0.0006 & 0.3673 & 0.0020 \\
8721.75426804 & 48.0785 & 1.0905 & 0.002300 & 0.2699 & 0.0004 & 0.3635 & 0.0019 \\
8722.74122029 & 17.4124 & 1.2063 & 0.029480 & 0.2744 & 0.0004 & 0.3701 & 0.0021 \\
8723.73967682 & -27.3998 & 1.0954 & 0.069110 & 0.2709 & 0.0004 & 0.3720 & 0.0022 \\
8724.82311116 & 3.5745 & 1.4927 & 0.030230 & 0.2679 & 0.0005 & 0.3718 & 0.0017 \\
8725.70980525 & 26.9388 & 1.1838 & 0.007080 & 0.2662 & 0.0004 & 0.3782 & 0.0017 \\
8734.72757798 & 27.6453 & 1.5650 & 0.011550 & 0.2669 & 0.0006 & 0.4008 & 0.0034 \\
8736.83306483 & 2.5551 & 1.9949 & 0.038890 & 0.2699 & 0.0006 & 0.3901 & 0.0047 \\
8737.80234571 & 11.0414 & 1.7889 & 0.034790 & 0.2792 & 0.0007 & 0.3674 & 0.0030 \\
8738.72461165 & 48.5519 & 1.8637 & -0.004540 & 0.2728 & 0.0006 & 0.3556 & 0.0025 \\
8739.74132231 & 9.4424 & 1.3690 & 0.046930 & 0.2720 & 0.0006 & 0.3488 & 0.0026 \\
8740.72127274 & 0.5538 & 1.5103 & 0.028690 & 0.2699 & 0.0006 & 0.3696 & 0.0027 \\
8741.71402357 & -8.6865 & 0.9042 & 0.064880 & 0.2677 & 0.0004 & 0.3569 & 0.0019 \\
8742.71298907 & 82.1742 & 1.3040 & 0.016950 & 0.2716 & 0.0005 & 0.3589 & 0.0019 \\
8744.86529394 & 12.8029 & 1.6896 & 0.010220 & 0.2738 & 0.0005 & 0.3616 & 0.0023 \\
8745.68610835 & -4.1308 & 2.2643 & 0.047710 & 0.2716 & 0.0008 & 0.3767 & 0.0024 \\
8746.72905288 & 32.9618 & 1.5638 & -0.011710 & 0.2759 & 0.0006 & 0.3596 & 0.0025 \\
8747.64660273 & 23.5623 & 2.8275 & 0.009380 & 0.2780 & 0.0010 & 0.3663 & 0.0025 \\
8748.66608518 & -2.7872 & 1.3049 & 0.043140 & 0.2731 & 0.0005 & 0.3500 & 0.0018 \\
\hline\hline
\end{longtable}
}


\section{Posteriors of the joint RV+TESS photometry fit}
\label{appendix:cornerplot}

\begin{figure*}[tb]
    \centering
    \begin{subfigure}[t]{\textwidth}
      \includegraphics[width=\textwidth]{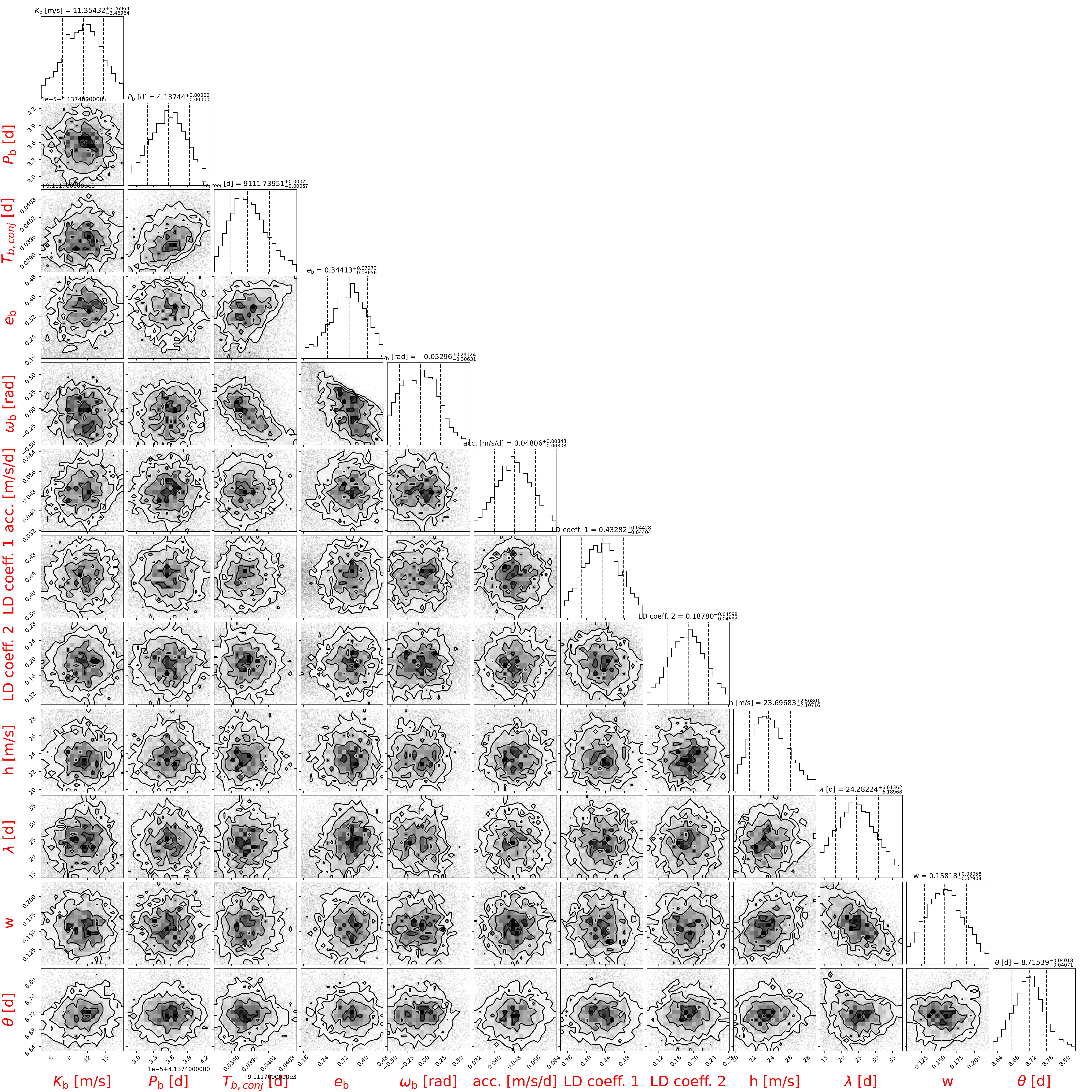}
    \end{subfigure}
    \caption{Posteriors of the free parameters used in the joint fit of the RV+TESS short-cadence light curve, including a linear long-term trend for the RVs (see Table \ref{t:toi-179_pl_param}). Two corner plots are shown for clarity.}
\end{figure*}

    \clearpage   

\begin{figure*}[tb]\ContinuedFloat
    \begin{subfigure}[t]{\textwidth}
      \includegraphics[width=\textwidth]{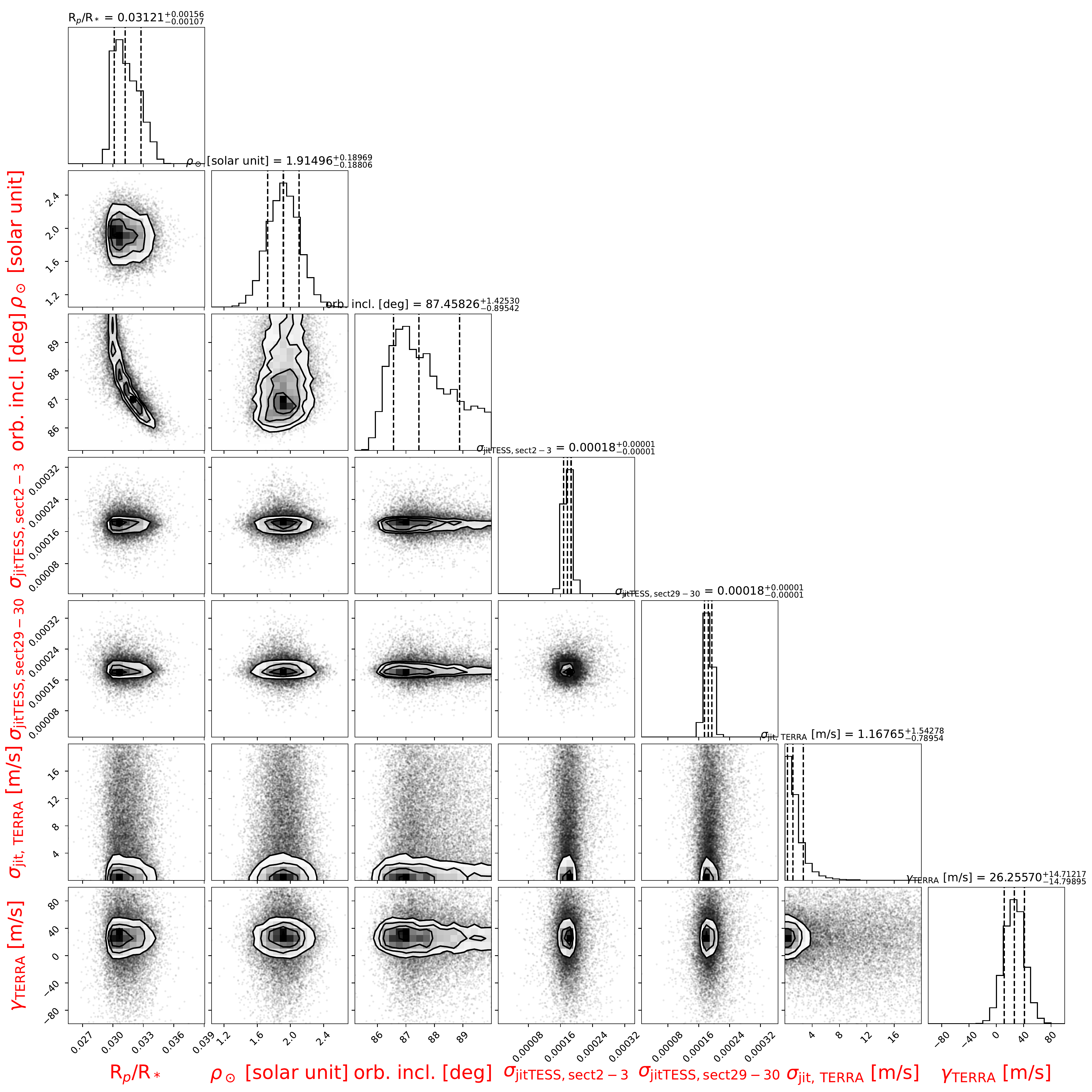}
    \end{subfigure}
    \caption{Continued}
    \label{fig:cornerplots}
\end{figure*}

    \end{document}